\RequirePackage{cmap}
\documentclass[%
	reprint,
superscriptaddress,
 amsmath,amssymb,
 aps,
 prstab,
]{revtex4-1}

\usepackage{graphicx}
\usepackage{dcolumn}
\usepackage{bm}
\usepackage{units}
\usepackage[colorlinks = true, linkcolor = blue,
            urlcolor  = blue,citecolor = blue,
            anchorcolor = blue]{hyperref}
\usepackage[noabbrev]{cleveref}
\usepackage[dvipsnames]{xcolor}

\renewcommand{\d}[2]{\frac{\mbox{d} #1}{\mbox{d} #2}} 

\begin{document}

\preprint{APS/123-QED}

\title{Characterisation of Laser Wakefield Acceleration Efficiency with Octave Spanning Near-IR Spectrum Measurements}


\author{M.J.V.~Streeter}
\email{m.streeter09@imperial.ac.uk}
\affiliation{The Cockcroft Institute, Keckwick Lane, Daresbury, WA4 4AD, United Kingdom}
\affiliation{Physics Department, Lancaster University, Lancaster LA1 4YB, United Kingdom}
\affiliation{The John Adams Institute for Accelerator Science, Imperial College London, London, SW7 2AZ, UK}

\author{Y.~Ma}
\affiliation{Center for Ultrafast Optical Science, University of Michigan, Ann Arbor, MI 48109-2099, USA}
\affiliation{The Cockcroft Institute, Keckwick Lane, Daresbury, WA4 4AD, United Kingdom}
\affiliation{Physics Department, Lancaster University, Lancaster LA1 4YB, United Kingdom}

\author{B.~Kettle}
\affiliation{The John Adams Institute for Accelerator Science, Imperial College London, London, SW7 2AZ, UK}

\author{S.J.D.~Dann}
\affiliation{The Cockcroft Institute, Keckwick Lane, Daresbury, WA4 4AD, United Kingdom}
\affiliation{Physics Department, Lancaster University, Lancaster LA1 4YB, United Kingdom}

\author{E.~Gerstmayr}
\affiliation{The John Adams Institute for Accelerator Science, Imperial College London, London, SW7 2AZ, UK}

\author{F.~Albert}
\affiliation{Lawrence Livermore National Laboratory (LLNL), P.O. Box 808, Livermore, California 94550, USA}

\author{N.~Bourgeois}
\affiliation{Central Laser Facility, STFC Rutherford Appleton Laboratory, Didcot OX11 0QX, UK}

\author{S.~Cipiccia}
\affiliation{Diamond Light Source, Harwell Science and Innovation Campus, Fermi Avenue, Didcot OX11 0DE, UK}

\author{J.M.~Cole}
\affiliation{The John Adams Institute for Accelerator Science, Imperial College London, London, SW7 2AZ, UK}

\author{I.~Gallardo~Gonz\'alez}
\affiliation{Department of Physics, Lund University, P.O. Box 118, S-22100, Lund, Sweden}


\author{A.E.~Hussein}
\affiliation{Center for Ultrafast Optical Science, University of Michigan, Ann Arbor, MI 48109-2099, USA}
\affiliation{Department of Electrical and Computer Engineering, University of Alberta, 9211 116 Street NW
Edmonton, Alberta, T6G 1H9, Canada}

\author{D.A.~Jaroszynski}
\affiliation{SUPA, Department of Physics, University of Strathclyde, Glasgow G4 0NG, UK}
\affiliation{The Cockcroft Institute, Keckwick Lane, Daresbury, WA4 4AD, United Kingdom}

\author{K.~Falk}
\affiliation{Helmholtz-Zentrum Dresden-Rossendorf, Bautzner Landstrasse 400, 01328 Dresden, Germany}
\affiliation{Technische Universit\:at Dresden, 01062, Dresden, Germany}
\affiliation{Institute of Physics of the ASCR, 182 21 Prague, Czech Republicy}

\author{K.~Krushelnick}
\affiliation{Center for Ultrafast Optical Science, University of Michigan, Ann Arbor, MI 48109-2099, USA}

\author{N.~Lemos}
\affiliation{Lawrence Livermore National Laboratory (LLNL), P.O. Box 808, Livermore, California 94550, USA}

\author{N.C.~Lopes}
\affiliation{The John Adams Institute for Accelerator Science, Imperial College London, London, SW7 2AZ, UK}
\affiliation{GoLP/Instituto de Plasmas e Fus\~{a}o Nuclear, Instituto Superior T\'{e}cnico, U.L., Lisboa 1049-001, Portugal}

\author{C.~Lumsdon}
\affiliation{York Plasma Institute, Department of Physics, University of York, York YO10 5DD, UK}

\author{O.~Lundh}
\affiliation{Department of Physics, Lund University, P.O. Box 118, S-22100, Lund, Sweden}

\author{S.P.D.~Mangles}
\affiliation{The John Adams Institute for Accelerator Science, Imperial College London, London, SW7 2AZ, UK}

\author{Z.~Najmudin}
\affiliation{The John Adams Institute for Accelerator Science, Imperial College London, London, SW7 2AZ, UK}

\author{P.P.~Rajeev}
\affiliation{Central Laser Facility, STFC Rutherford Appleton Laboratory, Didcot OX11 0QX, UK}

\author{R.~Sandberg}
\affiliation{Center for Ultrafast Optical Science, University of Michigan, Ann Arbor, MI 48109-2099, USA}

\author{M.~Shahzad}
\affiliation{SUPA, Department of Physics, University of Strathclyde, Glasgow G4 0NG, UK}
\affiliation{The Cockcroft Institute, Keckwick Lane, Daresbury, WA4 4AD, United Kingdom}

\author{M.~Smid}
\affiliation{Helmholtz-Zentrum Dresden-Rossendorf, Bautzner Landstrasse 400, 01328 Dresden, Germany}

\author{R.~Spesyvtsev}
\affiliation{SUPA, Department of Physics, University of Strathclyde, Glasgow G4 0NG, UK}
\affiliation{The Cockcroft Institute, Keckwick Lane, Daresbury, WA4 4AD, United Kingdom}

\author{D.R.~Symes}
\affiliation{Central Laser Facility, STFC Rutherford Appleton Laboratory, Didcot OX11 0QX, UK}

\author{G.~Vieux}
\affiliation{SUPA, Department of Physics, University of Strathclyde, Glasgow G4 0NG, UK}
\affiliation{The Cockcroft Institute, Keckwick Lane, Daresbury, WA4 4AD, United Kingdom}

\author{A.G.R.~Thomas}
\email{agrt@umich.edu}
\affiliation{Center for Ultrafast Optical Science, University of Michigan, Ann Arbor, MI 48109-2099, USA}
\affiliation{The Cockcroft Institute, Keckwick Lane, Daresbury, WA4 4AD, United Kingdom}
\affiliation{Physics Department, Lancaster University, Lancaster LA1 4YB, United Kingdom}

\date{\today}

\begin{abstract}
We report on experimental measurements of energy transfer efficiencies in a GeV-class laser wakefield accelerator. 
Both the transfer of energy from the laser to the plasma wakefield, and from the plasma to the accelerated electron beam were diagnosed by simultaneous measurement of the deceleration of laser photons and the acceleration of electrons as a function of plasma length. 
The extraction efficiency, which we define as the ratio of the energy gained by the electron beam to the energy lost by the self-guided laser mode, was maximised at $19\pm3$\% by tuning of the plasma density and length.  
The additional information provided by the octave-spanning laser spectrum measurement allows for independent optimisation of the plasma efficiency terms, which is required for the key goal of improving the overall efficiency of laser wakefield accelerators.
\end{abstract}

\maketitle

Intense laser pulses can drive compact plasma-based electron accelerators using a process known as Laser WakeField Acceleration (LWFA).
As the laser pulse propagates through a plasma, it drives electron oscillations that produce large electrostatic fields, typically of order \unit[100]{GV/m}. 
LWFA has been successfully used to accelerate electrons to \unit[$>1$]{GeV} energy levels over interaction distances on the order of a centimeter \cite{Leemans2006NP, Kneip2009PRL, Clayton2010PRL, Wang2013NC, Leemans2014PRL, Gonsalves2019PRL}.
A crucial consideration for LWFAs is the efficiency of energy transfer from the laser to the accelerated particle bunch.
In radio-frequency (RF) linear accelerators, efficient operation is achieved by storing the drive energy in a high quality-factor cavity, which is then extracted by multiple electron beams in a bunch train.
In high-amplitude plasma-accelerators non-linearities eventually damp out the plasma oscillations and so high efficiency energy transfer must be achieved within a relatively small number of plasma oscillation periods.

In plasma accelerators, the driver energy is converted to the accelerating fields via the plasma response so that the total efficiency of the accelerator can be broken down as $\eta = \left(\eta_{\mathrm{AC}\rightarrow\mathrm{driver}} \right) \cdot \left(\eta_{\mathrm{driver}\rightarrow\mathrm{plasma}} \right)\cdot \left(\eta_{\mathrm{plasma}\rightarrow\mathrm{beam}}\right)$, where the last term, from here on abbreviated as $\eta_b$, is the extraction efficiency.
In beam-driven plasma wakefield acceleration (PWFA), the extraction efficiency is simply calculated as the ratio of the energy gained by the witness beam, to the energy lost by the driver.
Using this measure, an efficiency of $>30$\% has been observed experimentally \cite{Litos2014N}.
In LWFA, energy transfer to the plasma wakefield occurs through redshifting of the driving laser pulse, and so can be determined from spectral measurement of the post-interaction laser pulse \cite{Esarey2009RMP,Shiraishi2013POP}.
Combined with measurement of the accelerated electron beam spectrum, it is possible to simultaneously diagnose the efficiency with which the laser excites the plasma wakefield, and the efficiency with which the electron beam extracts that energy. 
Higher-order laser modes that are not guided in a central filament will not drive strong plasma waves and therefore do not transfer significant energy to the wake \cite{Vieira2012PPCF}.
Consequently, the extraction efficiency for LWFA  only includes energy transfer from the guided, and therefore redshifted, laser mode.

Regardless of the nature of the driver, 100\% extraction efficiency would require that the wake of the witness beam perfectly cancels the plasma wake generated by the driver.
With a suitably chosen trapezoidal electron beam current profile \cite{Katsouleas1987}, the accelerating field over the electron bunch can be kept constant at $E_z(\xi_S)$ where $\xi_S$ is the location of the head of the electron bunch in the co-moving frame $\xi = z-ct$.
Doing so allows for simultaneous high plasma wake extraction efficiency and low energy spread for the accelerated beam in linear
\cite{Katsouleas1987} or non-linear blowout \cite{Gordienko2005POP,Lu2007PRSTAB,Tzoufras2008PRL} regimes.
If dephasing occurs, then this ideal beam-loading condition can not be maintained, leading to increased energy spread and lower overall efficiency.
Any modification to the wakefield amplitude, i.e. as the laser evolves, will affect both the ideal beam-loading condition and the dephasing rate.

In this letter, we present experimental measurements of the transfer of laser energy into a plasma wakefield and the efficiency with which that energy was extracted by an electron beam as it was accelerated to \unit[$>1$]{GeV}.
This required measurement of extended spectral range of the shifted laser pulse at the exit of the plasma, which in this regime extended up to \unit[1600]{nm}.
Studying the energy transfer between the laser, plasma and electron beam as a function of the plasma length was used to reveal the dynamics responsible for this optimum.

For the driving laser of a LWFA, assuming conservation of photon number (valid for $\Delta \omega < \omega_0$ and $\omega_p\ll \omega_0$ \cite{Bulanov1992PoFB,Silva1998PRE}, and negligible levels of ionization or incoherent scattering), the energy loss per unit length is given by $-\mathrm{d} W_L/\mathrm{d}z = -(W_{L0}/\omega_0)\mathrm{d}\langle\omega\rangle/\mathrm{d}z$, where $W_{L0}$ and $\omega_0$ are the  initial laser pulse energy and frequency, respectively. 
The electron-beam energy-gain per-unit-length, $N_B m_e c^2 d\langle\gamma\rangle/dz$, where $N_B$ is the number of accelerated electrons and $\langle\gamma\rangle = \int S(\gamma) \mathrm{d}\gamma/N_B$ is the average energy of the beam, can be inferred from measurements of the electron spectrum $S(\gamma)$ as a function of plasma length. 
Therefore, the instantaneous energy extraction efficiency for a LWFA can be written as,
\begin{equation}
\tilde{\eta_{b}} = -\frac{N_Bm_ec^2\omega_0}{W_{L0}}\left[\d{\langle\gamma\rangle}{z}\right]\left[\d{\langle\omega\rangle}{z}\right]^{-1} \;.
\label{eqn:efficencyMeasurement}
\end{equation}
Due to contributions by dephasing, drive laser evolution, beam injection and beam loss (changing $N_B$), the extraction efficiency is not a constant but changes along the accelerator length.
For the results of this paper, we measure the accelerator averaged extraction efficiency $\eta_b$, i.e. the ratio of the total energy gained by the electron beam to the energy lost by the laser pulse over the full acceleration length.

An experiment was performed (setup as shown in supplemental materials) with the Gemini laser at the Central Laser Facility.
Each pulse contained \unit[$6.3 \pm 0.6$]{J} in a pulse length of \unit[$\tau_{\rm FWHM} =52 \pm 4$]{fs}, with a peak power of \unit[$P_0 = 113 \pm 19$]{TW}. 
The pulse had a positive chirp of \unit[500]{fs$^2$} compared to the shortest pulse length of \unit[45]{fs}.
The pulse was focused to a spot width of \unit[$50 (\pm 2)$]{$\mu$m} \unit[$\times 40 (\pm 2)$]{$\mu$m} ($x \times y$ FWHM) using an $f/40$ parabolic mirror and was linearly polarised along the $x$-axis. 
A deformable mirror was used to optimize the wavefront, giving a peak intensity in vacuum of \unit[$I_0 = 3.9 (\pm 0.7 ) \times 10^{18}$]{W cm$^{-2}$} and a peak normalised vector potential $a_0=1.34\pm0.11$ at focus. 

The laser pulse was focused into a 3D printed two-stage gas cell \cite{Hussein2019SR}, filled with a 2\%/98\% nitrogen/helium mix for the first `injector' stage and pure helium in the second `accelerator' stage.
The cell walls had \unit[1]{mm} wide vertical slits to allow for the gas cell to be translated vertically.
This enabled the accelerator length to be adjusted continuously as the exit wall was angled at \unit[45]{$^{\circ}$} to the vertical plane.
The injector stage has an internal length of \unit[3]{mm} and the accelerator length was variable over \unit[8-21]{mm}, giving a total gas cell length \unit[14-27]{mm} (including the cell boundaries).
The electron density in the gas cells was varied in the range \unit[$n_e= 0-2.6 \times 10^{18}$]{cm$^{-3}$}, which was diagnosed by observing the spectrum of Raman side-scattering from plasma waves generated by low intensity ($a_0<1$), long duration (\unit[$\tau_{\mathrm{FWHM}}\approx 200$]{fs}) laser pulses \cite{Matsuoka2010PRL}.

After interaction with the plasma, the transmitted laser pulse was reflected from two glass plates into a fiber coupler, sampling a \unit[1]{cm} diameter region (1/10 of the full beam diameter at this point).
A fiber splitter directed the signal onto two spectrometers, one measuring \unit[350--840]{nm} (Andor Shamrock) and one measuring \unit[900--1700]{nm} (Ocean Optics NIRQuest 512).
The relative spectral sensitivities of the laser spectrometers was calibrated using a pre-calibrated white light source.
The electron beam spectrum was measured using a magnetic dipole with integrated field strength \unit[$\int B \mathrm{d}z=0.45$]{Tm}, which dispersed electrons in the energy range \unit[385--3000]{MeV} onto a Lanex scintillator.

\begin{figure*}[!ht] 
   \centering
   \includegraphics[width=17.9cm]{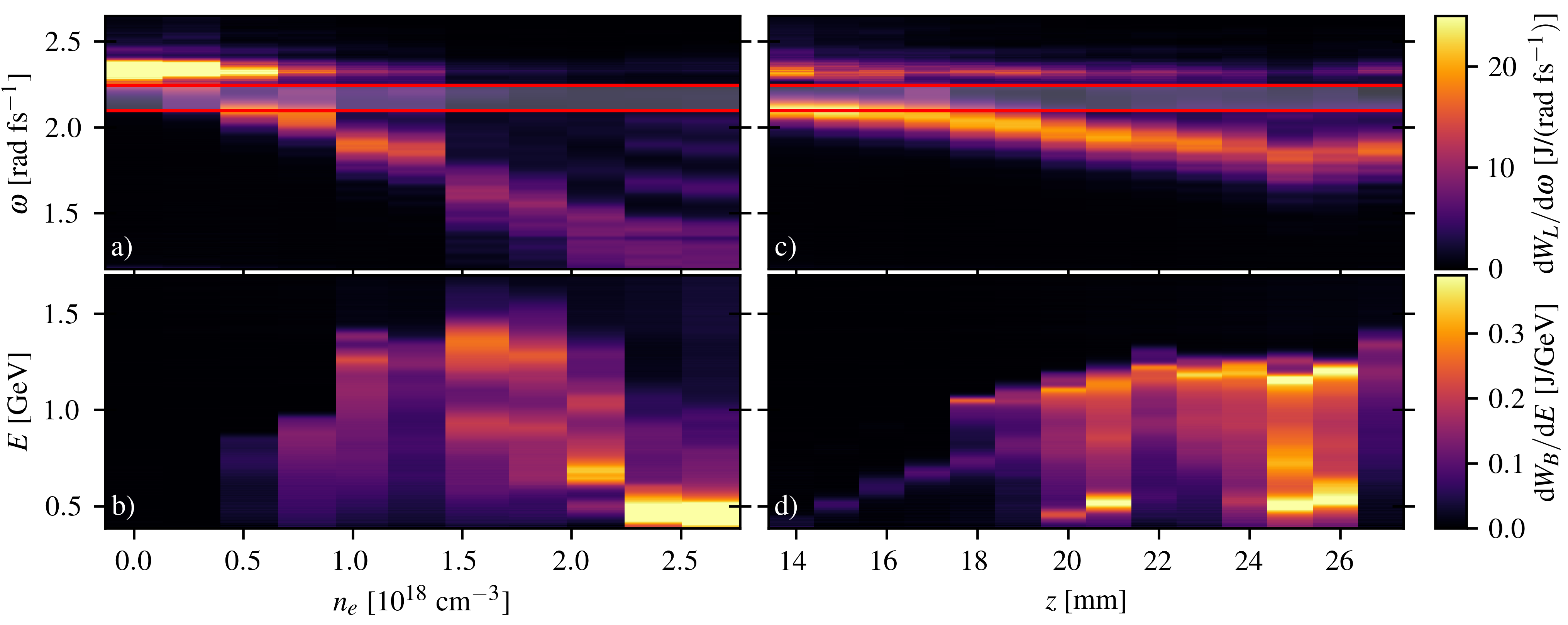} 
   \caption[Length scan]{ 
   Laser and electron spectrum at the exit of the \unit[27]{mm} gas cell as functions of electron density (a and b) and as functions of accelerator stage length for \unit[$n_e = 1.25 \pm 0.06 \times 10^{18}$]{cm$^{-3}$} (c and d).
   The red-bordered region in the laser spectra plots indicates the gap between the measurement ranges of the two spectrometers, which was filled by interpolation.
   Each column of each image shows the average from 3--5 shots at the same conditions.
   }
   \label{fig:expComp}
\end{figure*}

In order to determine the optimal conditions for electron generation, the gas cell was positioned at its longest length and the plasma density $n_e$ and longitudinal gas cell position were independently scanned.
The results of the gas cell density scan, for an accelerator length of \unit[21]{mm} (total plasma length \unit[27]{mm}), are plotted in \cref{fig:expComp}a-b.

The laser spectrum, shown in \cref{fig:expComp}a, was increasingly redshifted and broadened at higher plasma density.
For the highest plasma density \unit[$n_e = 2.6 \times 10^{18}$]{cm$^{-3}$}, the laser spectrum extended to the limit of the spectrometer with a peak occurring at \unit[1600]{nm}.
Comparatively little laser energy was blueshifted for the full density range demonstrating that ionization blueshift or photon acceleration at the rear of the plasma wave were not significant \cite{Schreiber2010PRL}.

As shown in \cref{fig:expComp}b, the highest electron energy occurred for \unit[$n_e = 1.5 \times 10^{18}$]{cm$^{-3}$}, where a peak in the spectrum was observed at \unit[$1.4$]{GeV}.
At higher densities, the maximum electron energies decreased, while the total measured charge remained approximately constant at \unit[$e N_B\approx210$]{pC}.
The increasing laser redshift indicates that a strong plasma wave continued to be driven at these high densities, but the injected electrons experienced less acceleration.

\Cref{fig:expComp}c-d show the result of scanning the accelerator length for a fixed density of of \unit[$n_e = 1.25\times 10^{18}$]{cm$^{-3}$}.
The laser (\cref{fig:expComp}c redshifted at an approximately linear rate as the acceleration length was increased.
The electron spectra (\cref{fig:expComp}d, shows two distinctive electron bunches were accelerated, with the higher energy component reaching \unit[1.2]{GeV}.
For plasma length \unit[$z<18$]{mm} (acceleration cell length \unit[$z<13$]{mm}) the charge in the higher energy component was much reduced, indicating that injection was sensitive to small changes in the plasma profile as the cell was translated vertically.
Also for the longest plasma lengths (\unit[$z>25$]{mm}), the laser redshift was reduced, indicating that less energy was coupled into the plasma. 
This was likely due to obstructions to the laser path at the top of the gas cell slits.

Both the laser and electron beam energies were calculated by integrating the measurements in \cref{fig:expComp} over the spectral axes.
For the laser spectrum, there was a small gap between the ranges of the two spectrometers.
In order to determine the interacting laser energy, Gaussian process regression (GPR) was used to fit the observed signal and interpolate over this region. 
The relative error of the interacting laser energy measurement introduced by this procedure was calculated from the GPR model uncertainties as less than $3$\% (standard deviation).
Furthermore, the spectra were corrected to account for frequency dependent divergence of the source, using the assumption that all frequencies were emitted from a constant spot size.
With this assumption, the divergence is inversely proportional to frequency and so the collection efficiency for an on-axis sample scales as $\omega^2$ and so the corrected spectrum was obtained by dividing the measured spectrum by $\omega^2$, i.e. $S_{\mathrm{cor.}}(\omega)  \propto S_{\mathrm{meas.}}(\omega)/ \omega^2$ .
Analysis of transmitted laser spectra from the PIC simulations (described later in this paper) indicate that this assumption leads to the laser depletion being underestimated by $\approx5$\%.
Finally, it was assumed that the total photon number of the driving laser pulse ($N_{ph} = W_{L0}/\hbar \omega_0$) was conserved throughout the interaction \cite{Esarey2009RMP}, such that the energy loss can be calculated from the change in average laser frequency as in \cref{eqn:efficencyMeasurement}.

\begin{figure}[!ht] 
  \centering
  \includegraphics[width=8.5cm]{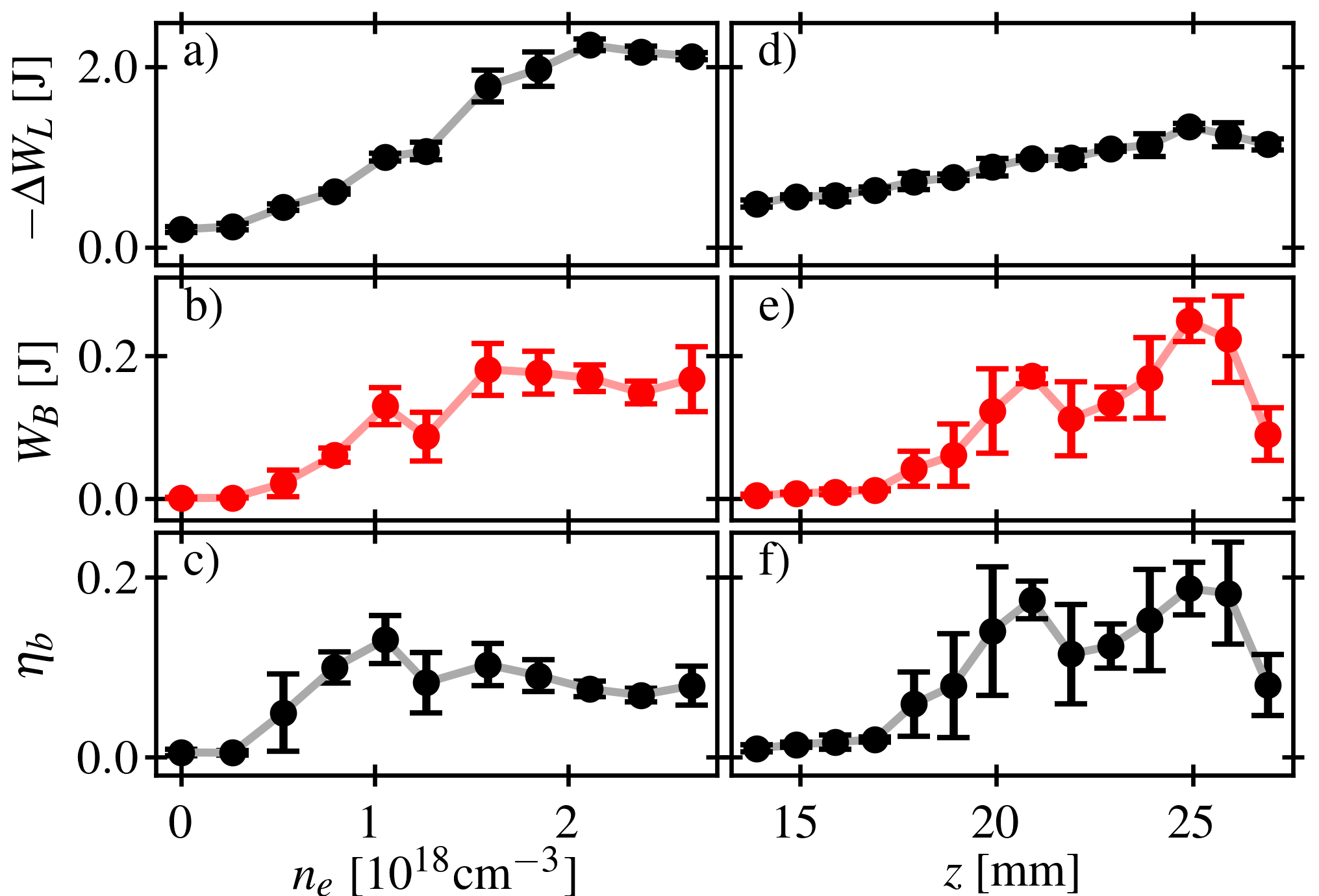} 
  \caption[Energy and efficiency]{Laser energy loss, electron beam energy and extraction efficiency as functions of plasma density for \unit[$z=27$]{mm} a-c), and length d-e) with \unit[$n_e=1.25\times10^{18}$]{cm$^{-3}$}.
  }
  \label{fig:expEfficiency}
\end{figure}

The laser pulse energy loss and electron energy gain are plotted as functions of plasma density in \cref{fig:expEfficiency}a \& b.
The laser lost more energy for increasing plasma density until reaching a plateau for \unit[$n_e > 1.6 \times 10^{18}$]{cm$^{-3}$}.
The electron beam total energy reached a maximum of \unit[$0.18\pm0.04$]{J} at \unit[$n_e = 1.6 \times 10^{18}$]{cm$^{-3}$} and was lower for both higher and lower plasma densities.
The extraction efficiency $\eta_b = -W_B/\Delta W_L$, plotted in \cref{fig:expEfficiency}c, reached a maximum of $13\pm3$\% at \unit[$n_e = 1.05 \times 10^{18}$]{cm$^{-3}$}.
Increasing the plasma density beyond this point resulted in a lower extraction efficiency, even though the electron beam energy increased. 

The laser energy loss and electron beam energy for the length scan are shown in \cref{fig:expEfficiency}d \& e.
The laser energy loss was approximately linearly proportional to the plasma length up to \unit[$z=25$]{mm}.
The electron beam total energy increased suddenly with the appearance of the higher energy feature in the electron spectrum (visible in \cref{fig:expComp}d), at \unit[$z=18$]{mm}.
Over the range \unit[$(20\leq z \leq 26)$]{mm}, the electron charge had an average of \unit[$eN_B=220\pm70$]{pC}, while the electron beam total energy increased with $z$.
The extraction efficiency, shown in \cref{fig:expEfficiency}f), also increased with length for \unit[$z<20$]{mm} as more charge was injected, before stabilizing at an average of $\eta_b = 16\pm3$\% for \unit[$(20\leq z \leq 26)$]{mm} with a maximum of $\eta_b = 19\pm3$\%.
The maximum electron beam total energy was \unit[$W_b=0.25 \pm 0.03$]{J}, giving a total LWFA efficiency $\eta_{\mathrm{laser \rightarrow beam}} = 4.0 \pm 0.5\%$.

In order to explore the dynamics of the experiment, we performed quasi-3D PIC simulations using FBPIC \cite{Lehe2016CPC}, using 4 azimuthal modes (see details in supplemental material).
The laser pulse was initialised using the experimentally measured temporal profile, and a Gaussian approximation to the measured focal spot energy distribution with an $a_0=1.34$.
The simulation results matched the experimentally observed maximum electron energy and laser redshift, with relative differences of 5\% and 0.1\% respectively.
However, the simulated idealised laser pulse contained only \unit[4.2]{J} of energy (\unit[$66$]{\%} of the laser energy in the experiment), indicating that the combination of laser pulse and target imperfections resulted in a lower proportion of the laser energy being guided than for a pure Gaussian mode, in line with previous observations \cite{Mangles2012PRSTAB,Vieira2012PPCF}.

\begin{figure}[!ht] 
   \centering
   \includegraphics[width=8.5cm]{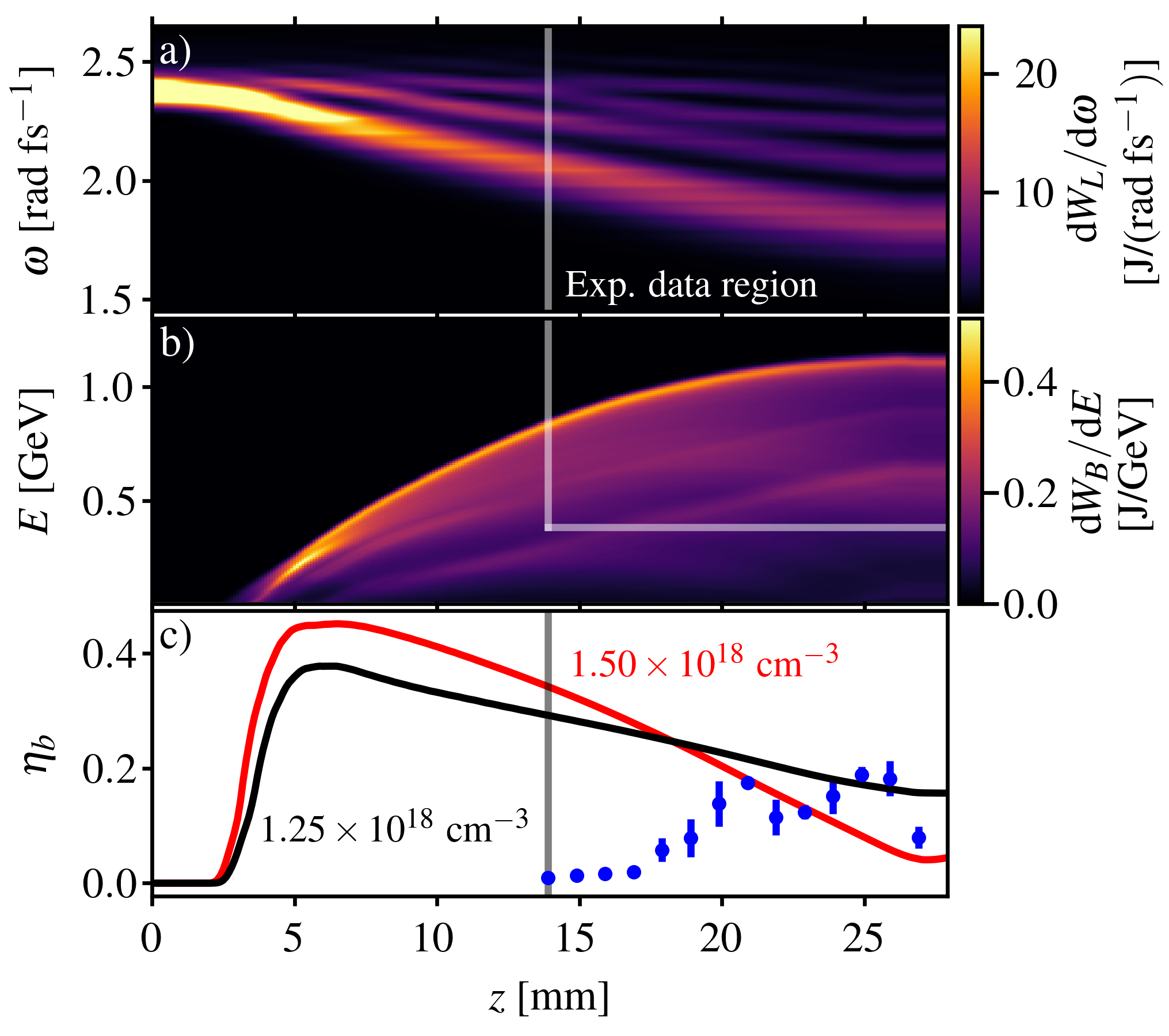} 
   \caption[Simulation redshift]{a) Laser and b) electron spectra as functions of propagation distance taken from a PIC simulation with a plateau density of \unit[$n_e = 1.25\times10^{18}$]{cm$^{-3}$}.
   c) The extraction efficiency, calculated as the ratio of the electron beam energy to the laser energy loss, for simulations with \unit[$n_e = (1.25,1.5)\times10^{18}$]{cm$^{-3}$} and the experimental measurements (blue points) at \unit[$n_e = 1.25\times10^{18}$]{cm$^{-3}$}.
   }
   \label{fig:sim_spectra}
\end{figure}

The laser and electron spectra, as functions of propagation distance within the PIC simulation, are shown in \cref{fig:sim_spectra}a and b.
The laser spectrum was seen to redshift and broaden by the same amount as the experimental measurements of \cref{fig:expComp}c, with a redshifted peak at \unit[$\omega=1.9$]{rad fs$^{-1}$}.
The laser reached a peak normalised vector potential of $a_0=3$ at \unit[$z=4.4$]{mm}, due to the effects of self-focusing and self-compression.
For the rest of the accelerator, $a_0<3$ and no self-injection was observed.
The accelerated electron beam resulted from trapping of inner shell electrons from nitrogen (ionisation injection \cite{RowlandsRees2008PRL,Pak2010PRL,McGuffey2010PRL,Chen2012POP}) during the first \unit[5]{mm}.
A total beam charge of \unit[255]{pC} was observed in the accelerated beam at the end of the simulation, with a peak in energy of \unit[1.2]{GeV}.
Approximately 16\% of the electron beam energy at the end of the simulation lies below the \unit[385]{MeV} detection threshold of the experimental spectrometer, indicating that the experimentally measured extraction efficiency is underestimated by a similar amount.

The extraction efficiency is shown in \Cref{fig:sim_spectra}c for \unit[$n_e = 1.25\times10^{18}$]{cm$^{-3}$} and for a slightly higher density \unit[$n_e = 1.5\times10^{18}$]{cm$^{-3}$}.
For the higher density case, the laser intensity reached a larger value during the initial self-focusing, resulting in the trapping of $\sim40$\% more charge.
Also for the higher density, the laser pulse maintained $a_0>2.5$ over the full acceleration distance, thereby driving a higher amplitude wake.
As a result, beam-loading was less severe and so the extraction efficiency was initially higher than for the lower density case.
However, for the higher density case the efficiency dropped significantly throughout the accelerator, as the electron beam dephased and started to decelerate for \unit[$z>17$]{mm}.

Simulations using the same plasma profile with plateau densities of \unit[$n_e =(1.1,1.2,1.3,1.4)\times 10^{18}$]{cm$^{-3}$}, show that both the maximum final electron energy and extraction efficiencies were optimised for\unit[$n_e =1.2\times 10^{18}$]{cm$^{-3}$}.
Operating the LWFA at a higher density but over a shorter length allows for higher efficiency, but results in a lower maximum electron energy.

\begin{figure}[!t] 
   \centering
   \includegraphics[width=8.5cm]{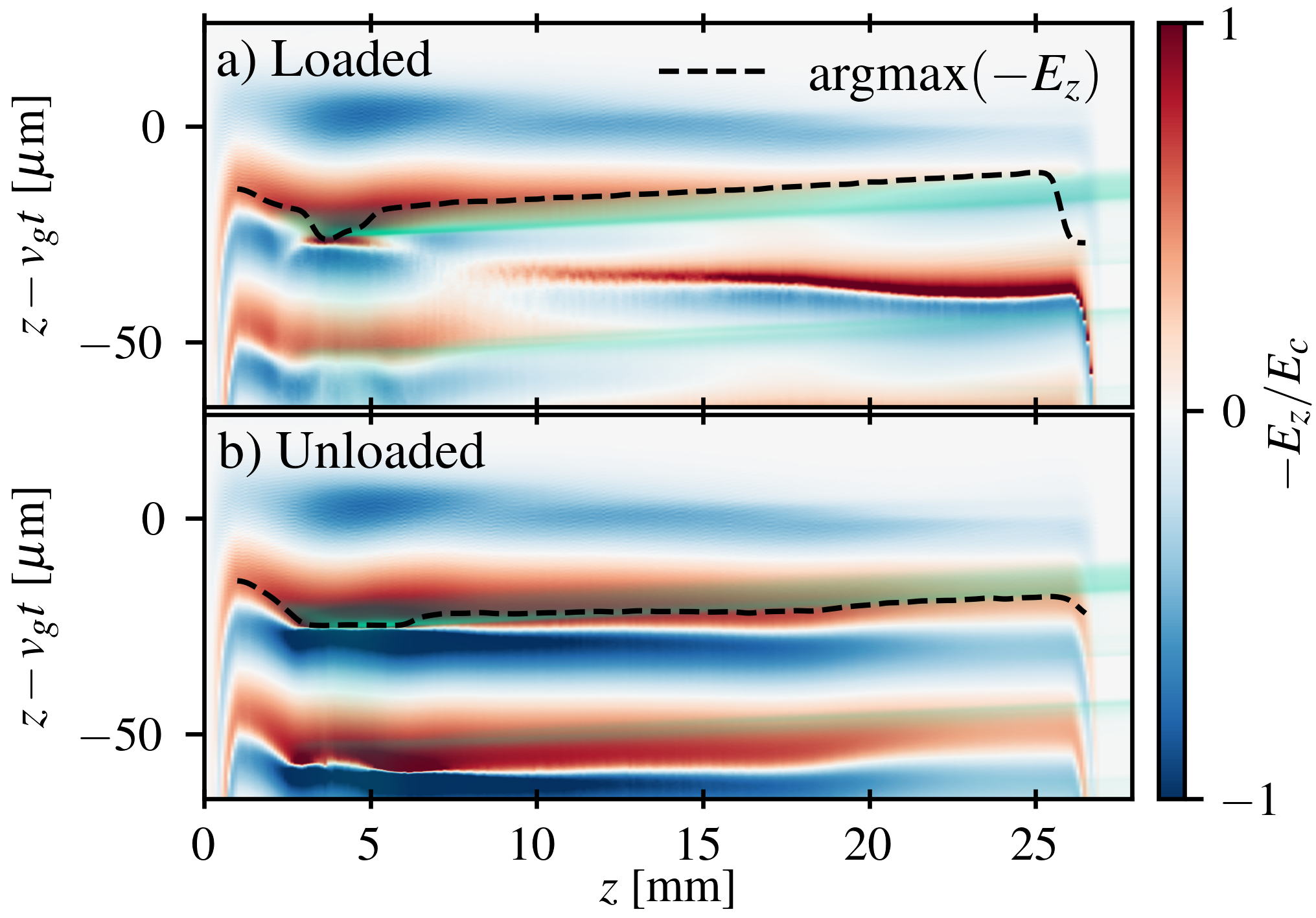} 
   \caption[Simulation redshift]{a) Loaded and b) unloaded longitudinal electric field and injected electron bunch position (green) over the propagation axis $z$ of PIC simulations with \unit[$n_e = 1.25\times10^{18}$]{cm$^{-3}$}.
   The unloaded fields were extracted from a simulation without the ionization injection species.
   The maximum accelerating field position in the first plasma period (black line) is overlaid.
   }
   \label{fig:simBeamField}
\end{figure}

\Cref{fig:simBeamField} shows the electron beam position relative to the axial longitudinal electric field in the lab frame with a Galilean coordinate transform to the linear group velocity of the laser $\xi = z - v_gt$.
The fields are shown both with (\cref{fig:simBeamField}a) and without (\cref{fig:simBeamField}b) the contribution of the trapped electron bunch.
After the initial self-focusing phase, the laser pulse $a_0$ steadily dropped from $a_0=2$ to $a_0=1.5$ at \unit[$z=22$]{mm}, causing a gradual reduction in the wavelength of the wakefield $\lambda_p$.
In addition, as the shape of the wakefield smoothly changed from the saw-tooth profile of the highly non-linear regime to a quasi-linear sinusoidal profile, which also had the effect of moving the position of peak accelerating field further forward.
As shown in \cref{fig:simBeamField}b, these effects combined to balance the subluminal group velocity of the driving laser pulse and so the trapped highly relativistic electrons closely tracked the position of highest field strength.
\Cref{fig:simBeamField}a, shows the effect of beam-loading, which reduced the accelerating field experienced by the rear of the bunch to $\approx15$\% of the value at the head.

During injection, the normalised vector potential reached a peak value of $a_0=3$, significantly above the threshold for ionisation injection.
Electrons injected at this point were trapped on \emph{deep} orbits \cite{Chen2012POP} in which electrons obtained the wakefield phase velocity $v=v_{\phi}$ significantly before the back of the wake, and so never experienced the maximum accelerating wakefield.
From this point on, a non-evolving laser driver would have caused the electron beam to dephase at \unit[$L_\phi=14.5$]{mm} with a maximum energy of \unit[800]{MeV}.
However, the driver evolution acted to mitigate dephasing, resulting in the higher observed electron beam energy of \unit[1.2]{GeV}.

Methods for accelerating electrons to energies beyond the dephasing limit have been explored, including non-uniform plasma profiles \cite{Pukhov2008PRE,Guillaume2015PRL, Ma2018POP,Sadler2020PRAB} or by using alternative laser focusing geometries with spatio-temporal couplings \cite{Debus2019PRX,Palastro2020PRL,Caizergues2020Nph}.
Phase-locked LWFA dynamics in a constant density plasma have been observed in PIC simulations previously \cite{Li2014APL} although in that case it was attributed to pulse depletion.
However, depletion increases the wakefield amplitude due to laser redshifting ($a_0 \propto \omega^{-1/2}$) and so this effect alone would actually increase the wakefield wavelength and cause the electron beam to dephase more rapidly.
Here we show that through careful management of the laser evolution it is possible to mitigate dephasing, so that pulse depletion determines the electron energy limit for the accelerator.

In conclusion, we have measured the extraction efficiency for an LWFA, which reached a maximum of $19\pm3$\%, close to that previously observed in electron-beam-driven PWFA \cite{Litos2014N}. 
The measurements indicated that only approximately 20\% of the laser pulse energy was transferred to the plasma wakefield, with approximately one third of the laser energy wasted due to a non-ideal focal spot.
The overall efficiency could therefore be increased by improving the spatial distribution of the laser pulse, as indicated by PIC simulations with a Gaussian distribution as well as previous studies on the effects of non-Gaussian focal spots \cite{Vieira2012PPCF}.
Laser energy that remains after the interaction could possibly be recovered to further improve the total efficiency of a LWFA facility.
For the highest plasma densities and longest interaction length, the laser pulse spectrum was observed to span a complete octave from \unit[800--1600]{nm}.
Further harnessing of these effects may open up a route to relativistic intensity single cycled mid-IR laser pulses \cite{Streeter2018PRL,Nie2018Nph,Nie2020NC}.

The data and analysis scripts are available at the online
repository zenodo.org at \cite{Streeter2022zenodo}.

\begin{acknowledgments}
We acknowledge support from the UK STFC core Grants No. ST/P002056/1 (Cockcroft Institute), No. ST/P000835/1, No. ST/P002021/1, and No. ST/V001639/1 (John Adams Institute), UK EPSRC (EP/J018171/1, EP/N028694/1), the European Union’s Horizon 2020 research and innovation programme under grant agreement Laserlab-Europe (871124), the National Science Foundation (Grant No. 1804463) and the Air Force Office of Scientific Research (Grant No. FA9550-16-1-0121). F. A. acknowledges funding from the DOE Early Career research program (Fusion Energy Sciences SCW1575-1).
\end{acknowledgments}

\section*{Supplemental}

\subsection*{Experimental Setup}

\begin{figure}[!ht] 
 \centering
 \includegraphics[width=8.5cm]{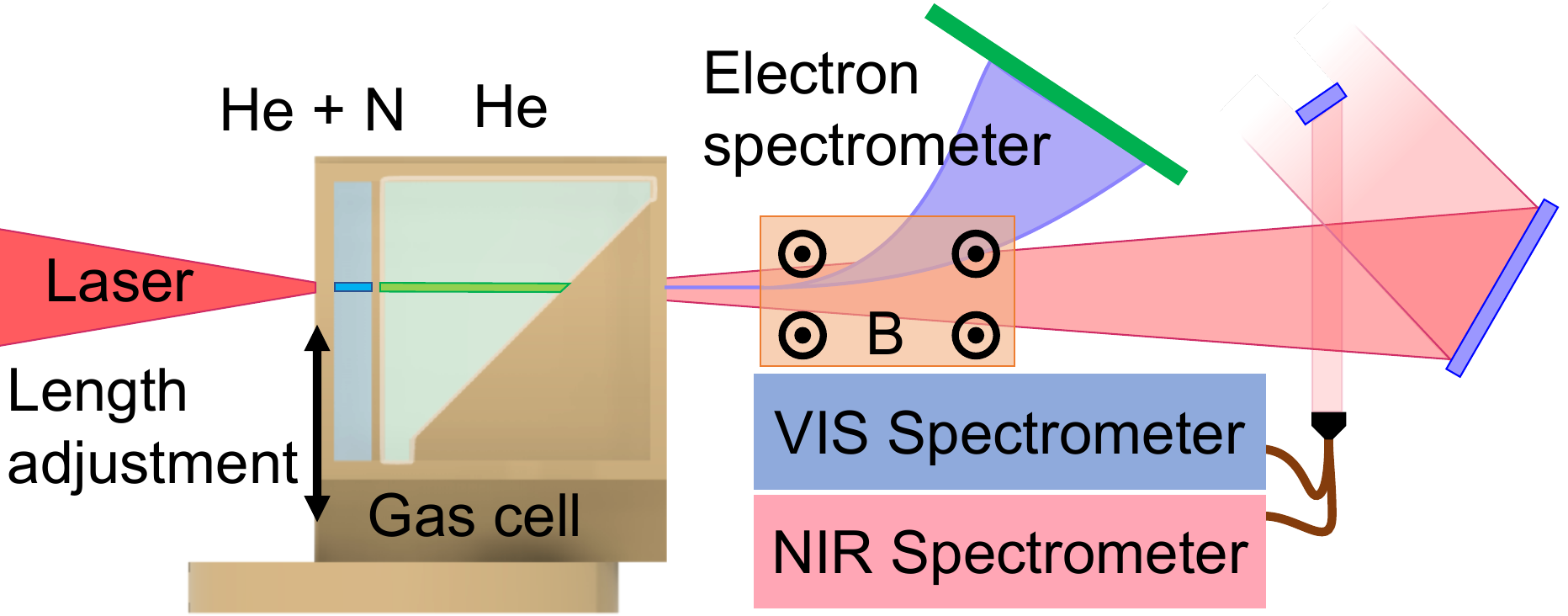} 
 \caption[Experimental Setup]{Experimental schematic showing the gas cell geometry, the electron spectrometer and the NIR (\unit[900--1700]{nm}) and VIS (\unit[350--840]{nm}) spectrometers which measured the transmitted laser pulse.
 The input laser is horizontally polarized, whereas the dispersion axis of the electron spectrometer is in vertical plane.
 The transmitted laser is reflected in the horizontal plane by two glass plates to collected the central region.}
 \label{fig:expSetup}
\end{figure}

\Cref{fig:expSetup} shows a conceptual sketch of the experimental setup.
The the length of the acceleration stage of the gas cell was varied by translating the gas cell vertically.
The longest gas cell (\unit[27]{mm} total length) was achieved by aligning the laser to the top of the entrance and exit slits.
A dividing wall with a vertical slit was used to confine gas to the two compartments of the gas cell.
Scattered laser light was imaged through the glass slides at the two sides of the gas cell.

\begin{figure}[!ht] 
 \centering
 \includegraphics[width=8.5cm]{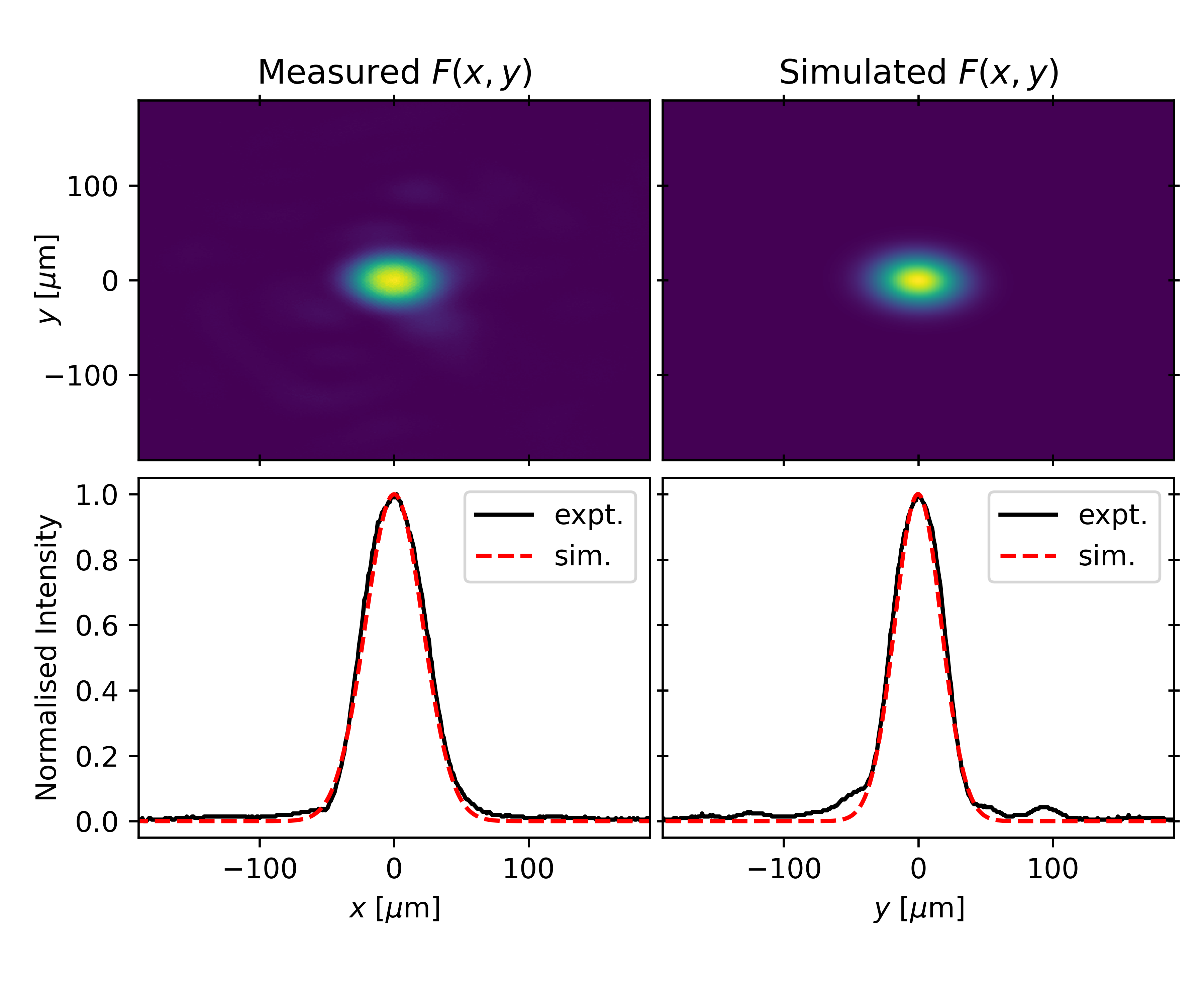} 
 \caption[Focal spots]{Measured (left) and fitted (right) normalised focal spot. A 2D Gaussian fit was performed which gave a FWHM focal spot width of \unit[$50\times40$]{$\mu$m}.}
 \label{fig:focalSpot}
\end{figure}

The laser focus was measured with a $\times 10$ microscope objective which imaged the focal plane onto a CCD. A 2D Gaussian fit was performed to obtain an analytical approximation to the laser for particle-in-cell simulations.
\Cref{fig:focalSpot} shows an example measured and fitted focal spot image.

\begin{figure}[!ht] 
 \centering
 \includegraphics[width=10cm]{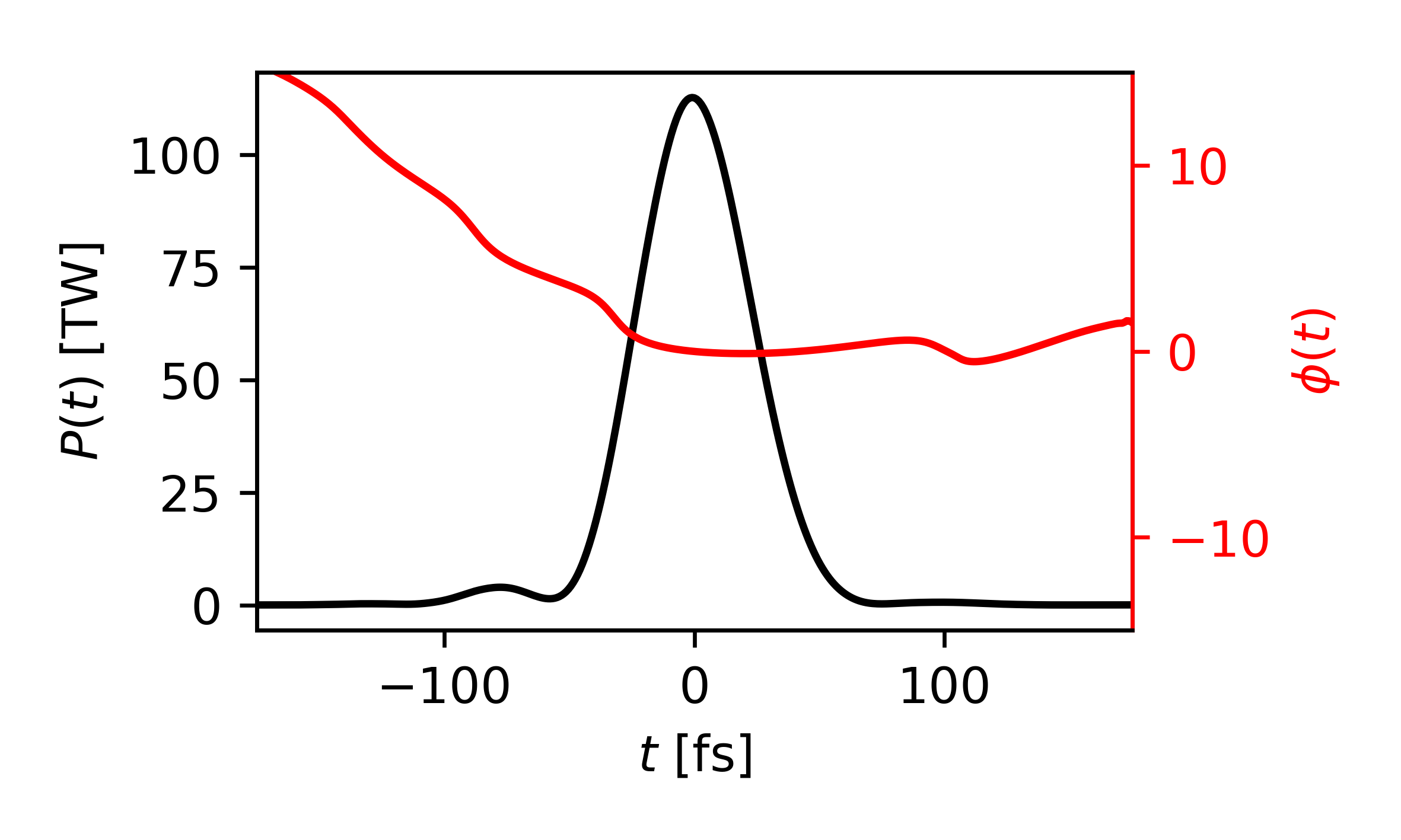} 
 \caption[Pulse profile]{The laser pulse temporal profile, as measured by a SHG FROG. The same temporal profile was using in the particle-in-cell simulations.}
 \label{fig:pulseProfile}
\end{figure}

The temporal profile of the laser pulse, as measured by a second harmonic generation frequency resolved optical gating (SHG-FROG) is show in \cref{fig:pulseProfile}. 
The sample profile was using in the particle-in-cell simulations.

\subsection*{Experimental data}

\begin{figure*}[!ht] 
 \centering
 \includegraphics[width=16cm]{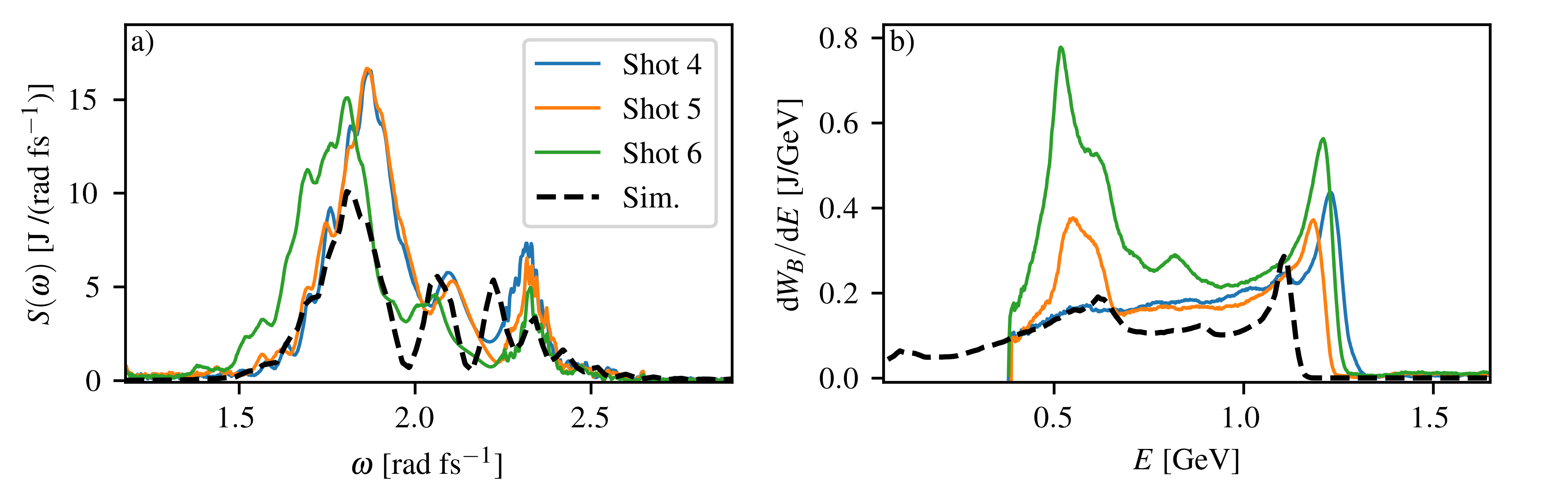} 
 \caption[Laser spectra]{Example a) post-plasma laser spectra and b) electron spectra for \unit[$n_e=1.25\times10^{18}$]{cm$^{-3}$} and \unit[$z=25$]{mm}.
 Each individual shot is shown for the experimental data.}
 \label{fig:ExampleSpectra}
\end{figure*}

\begin{figure*}[!ht] 
 \centering
 \includegraphics[width=16cm]{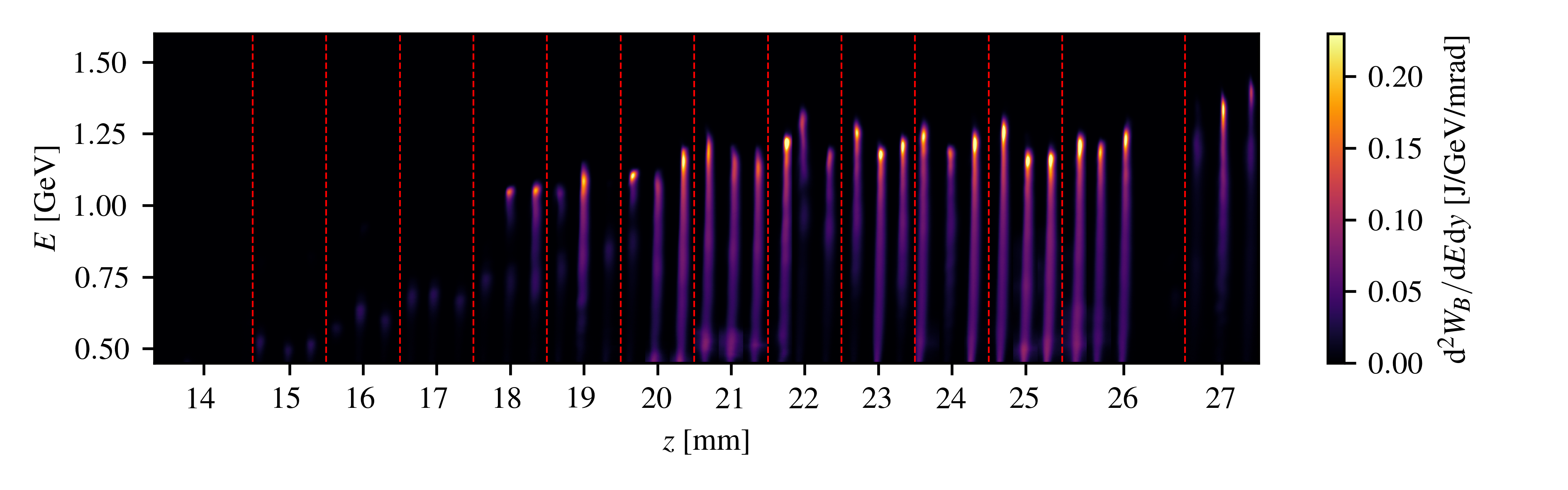} 
 \caption[Electron montage]{A montage of the electron spectra from the length scan at \unit[$n_e=1.25\times10^{18}$]{cm$^{-3}$}.
 Each spectrum is shown for the lengths given in the $x$-axis.}
 \label{fig:electronMontage}
\end{figure*}

Example electron and laser spectra from three shots with identical conditions are are shown in \cref{fig:ExampleSpectra}.
Also shown in \cref{fig:electronMontage} is a montage of electron beams produced in the length-scan at \unit[$n_e=1.25\times10^{18}$]{cm$^{-3}$}.

\subsubsection*{Interpolation across gap in laser spectrum measurement}

\begin{figure}[!ht] 
 \centering
 \includegraphics[width=8.5cm]{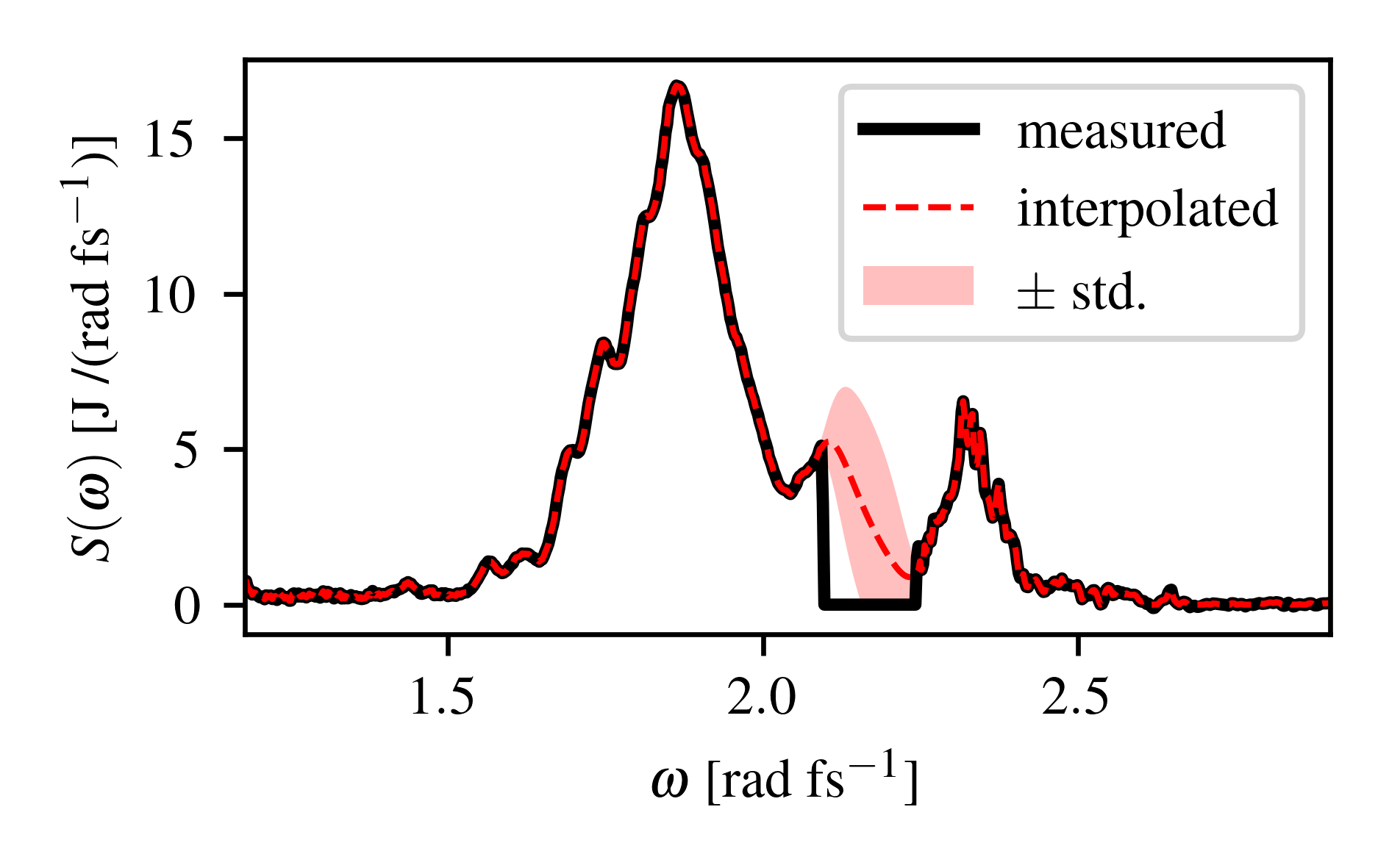} 
 \caption[Laser interpolation]{The measured laser spectrum along with the interpolated signal and its uncertainty using a Gaussian process regression model for a single shot with \unit[$n_e=1.25\times10^{18}$]{cm$^{-3}$} and \unit[$z=26$]{mm}.}
 \label{fig:exampleLaserSpectrum}
\end{figure}

Due to the gap in the laser spectrum measurement, interpolation was required to more accurately measure the average frequency of the transmitted laser pulse. 
This was performed by training a Gaussian-process regression model on the measured region of the spectrum and using it to predict the signal in the measurement gap. 
A new model was trained for each shot, using 200 randomly selected points from the measured spectrum and a Matern kernel with a fixed scale length of \unit[0.05]{rad fs$^{-1}$}. 
The relative error on the inferred laser energy due the uncertainty of this interpolation was estimated by taking the standard deviation of the laser energy using 100 random samples from the Gaussian process model. 
\Cref{fig:exampleLaserSpectrum} shows the results of an interpolation.

\subsubsection*{Measured photon number vs plasma length}
\begin{figure}[!ht] 
 \centering
 \includegraphics[width=8.5cm]{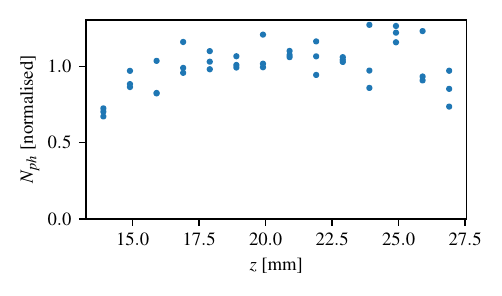} 
 \caption[Laser interpolation]{The relative measured photon number as a function of plasma length for a plasma density of \unit[$n_e=1.25\times10^{18}$]{cm$^{-3}$}.}
 \label{fig:photonNumber}
\end{figure}

The measurement of the laser energy from the average frequency relies on the assumption that the number of photons is constant. 
This assumption has a strong justification based on the fundamentals of the forward Raman-scattering process that drives the plasma wave in laser-wakefield acceleration.
The average frequency measurement is much more robust to changes in collection efficiency and allows for on-axis sampling to give an accurate measurement of the transmitted laser energy.
An approximate test of this assumption can be performed by measuring the photon number of the measured laser spectrum, under the assumption that the sampling collection efficiency scales with $\omega^2$.
This is plotted as a function of the plasma cell length in \cref{fig:photonNumber}.

\subsection*{Simulations}

Simulations were performed using the quasi-3D Particle-In-Cell (PIC) code FBPIC 
(\url{https://fbpic.github.io/index.html}) using cylindrical symmetry with azimuthal mode decomposition.
The simulation domain was \unit[$95 \times 85$]{$\mu$m} in the propagation ($z$) and radial $r$ axes, which were divided into \unit[$N_z \times N_r = 1900 \times 240$]{cells} with four azimuthal modes ($N_m=4$).
The simulation window co-propagated with the laser, with the window velocity set equal to the linear group velocity of the laser $\beta_g = \sqrt{1-\omega_p^2/\omega_0^2}$.
The simulation time-step was \unit[0.167]{fs}.
The plasma electrons, helium ions and nitrogen ions were each represented by $2\times2\times16$ macro-particles per cell in the $z$, $r$ and $\theta$ directions.
Open boundary conditions were used for both the $z$ and $r$ directions.

The plasma was initialized with a density of \unit[$1.25 \times 10^{18}$]{mm$^{-3}$} (once the helium was fully ionised), with \unit[$500$]{$\mu$m} linear ramps either side of a \unit[26]{mm} plateau.
The helium ions were initalised as singly ionised and nitrogen was initially ionised up the the 5th level.
The ionisation thresholds were calculated using the Ammosov-Delone-Krainov (ADK) model.
The temporal profile of the laser pulse was taken from FROG measurements of the input laser pulses during the experiment.
Measurements of the laser focal spot were used to fit a 2D Gaussian transverse profile as $a(x,y,t) = \hat{a}(t) \exp\left[-(x/\sigma_x)^2-(y/\sigma_y)^2\right]$ with \unit[$\sigma_x = 40$]{$\mu$m} and \unit[$\sigma_y = 32$]{$\mu$m}.
The laser was linearly polarized in the $x$-direction.

Convergence testing was performed to examine the effects of the transverse resolution, by varying the number of transverse cells $N_r$ and the number of azimuthal modes $N_m$.
Otherwise identical simulations were performed for $N_r=100,200,300$ and $N_m=2,3,4$.
No appreciable difference was seen in the laser spectrum, electron beam energy, transfer efficiency or the longitudinal electric field as these parameters were changed, indicating that they were not significantly affected by numerical noise.


\begin{thebibliography}{37}%
   \makeatletter
   \providecommand \@ifxundefined [1]{%
    \@ifx{#1\undefined}
   }%
   \providecommand \@ifnum [1]{%
    \ifnum #1\expandafter \@firstoftwo
    \else \expandafter \@secondoftwo
    \fi
   }%
   \providecommand \@ifx [1]{%
    \ifx #1\expandafter \@firstoftwo
    \else \expandafter \@secondoftwo
    \fi
   }%
   \providecommand \natexlab [1]{#1}%
   \providecommand \enquote  [1]{``#1''}%
   \providecommand \bibnamefont  [1]{#1}%
   \providecommand \bibfnamefont [1]{#1}%
   \providecommand \citenamefont [1]{#1}%
   \providecommand \href@noop [0]{\@secondoftwo}%
   \providecommand \href [0]{\begingroup \@sanitize@url \@href}%
   \providecommand \@href[1]{\@@startlink{#1}\@@href}%
   \providecommand \@@href[1]{\endgroup#1\@@endlink}%
   \providecommand \@sanitize@url [0]{\catcode `\\12\catcode `\$12\catcode
     `\&12\catcode `\#12\catcode `\^12\catcode `\_12\catcode `\%12\relax}%
   \providecommand \@@startlink[1]{}%
   \providecommand \@@endlink[0]{}%
   \providecommand \url  [0]{\begingroup\@sanitize@url \@url }%
   \providecommand \@url [1]{\endgroup\@href {#1}{\urlprefix }}%
   \providecommand \urlprefix  [0]{URL }%
   \providecommand \Eprint [0]{\href }%
   \providecommand \doibase [0]{http://dx.doi.org/}%
   \providecommand \selectlanguage [0]{\@gobble}%
   \providecommand \bibinfo  [0]{\@secondoftwo}%
   \providecommand \bibfield  [0]{\@secondoftwo}%
   \providecommand \translation [1]{[#1]}%
   \providecommand \BibitemOpen [0]{}%
   \providecommand \bibitemStop [0]{}%
   \providecommand \bibitemNoStop [0]{.\EOS\space}%
   \providecommand \EOS [0]{\spacefactor3000\relax}%
   \providecommand \BibitemShut  [1]{\csname bibitem#1\endcsname}%
   \let\auto@bib@innerbib\@empty
   \bibitem [{\citenamefont {Leemans}\ \emph {et~al.}(2006)\citenamefont
     {Leemans}, \citenamefont {Nagler}, \citenamefont {Gonsalves}, \citenamefont
     {T{\'{o}}th}, \citenamefont {Nakamura}, \citenamefont {Geddes}, \citenamefont
     {Esarey}, \citenamefont {Schroeder},\ and\ \citenamefont
     {Hooker}}]{Leemans2006NP}%
     \BibitemOpen
     \bibfield  {author} {\bibinfo {author} {\bibfnamefont {W.~P.}\ \bibnamefont
     {Leemans}}, \bibinfo {author} {\bibfnamefont {B.}~\bibnamefont {Nagler}},
     \bibinfo {author} {\bibfnamefont {A.~J.}\ \bibnamefont {Gonsalves}}, \bibinfo
     {author} {\bibfnamefont {C.}~\bibnamefont {T{\'{o}}th}}, \bibinfo {author}
     {\bibfnamefont {K.}~\bibnamefont {Nakamura}}, \bibinfo {author}
     {\bibfnamefont {C.~G.~R.}\ \bibnamefont {Geddes}}, \bibinfo {author}
     {\bibfnamefont {E.}~\bibnamefont {Esarey}}, \bibinfo {author} {\bibfnamefont
     {C.~B.}\ \bibnamefont {Schroeder}}, \ and\ \bibinfo {author} {\bibfnamefont
     {S.~M.}\ \bibnamefont {Hooker}},\ }\href {\doibase 10.1038/nphys418}
     {\bibfield  {journal} {\bibinfo  {journal} {Nature Physics}\ }\textbf
     {\bibinfo {volume} {2}},\ \bibinfo {pages} {696} (\bibinfo {year}
     {2006})}\BibitemShut {NoStop}%
   \bibitem [{\citenamefont {Kneip}\ \emph {et~al.}(2009)\citenamefont {Kneip},
     \citenamefont {Nagel}, \citenamefont {Martins}, \citenamefont {Mangles},
     \citenamefont {Bellei}, \citenamefont {Chekhlov}, \citenamefont {Clarke},
     \citenamefont {Delerue}, \citenamefont {Divall}, \citenamefont {Doucas},
     \citenamefont {Ertel}, \citenamefont {Fiuza}, \citenamefont {Fonseca},
     \citenamefont {Foster}, \citenamefont {Hawkes}, \citenamefont {Hooker},
     \citenamefont {Krushelnick}, \citenamefont {Mori}, \citenamefont {Palmer},
     \citenamefont {Phuoc}, \citenamefont {Rajeev}, \citenamefont {Schreiber},
     \citenamefont {Streeter}, \citenamefont {Urner}, \citenamefont {Vieira},
     \citenamefont {Silva},\ and\ \citenamefont {Najmudin}}]{Kneip2009PRL}%
     \BibitemOpen
     \bibfield  {author} {\bibinfo {author} {\bibfnamefont {S.}~\bibnamefont
     {Kneip}}, \bibinfo {author} {\bibfnamefont {S.~R.}\ \bibnamefont {Nagel}},
     \bibinfo {author} {\bibfnamefont {S.~F.}\ \bibnamefont {Martins}}, \bibinfo
     {author} {\bibfnamefont {S.~P.~D.}\ \bibnamefont {Mangles}}, \bibinfo
     {author} {\bibfnamefont {C.}~\bibnamefont {Bellei}}, \bibinfo {author}
     {\bibfnamefont {O.}~\bibnamefont {Chekhlov}}, \bibinfo {author}
     {\bibfnamefont {R.~J.}\ \bibnamefont {Clarke}}, \bibinfo {author}
     {\bibfnamefont {N.}~\bibnamefont {Delerue}}, \bibinfo {author} {\bibfnamefont
     {E.~J.}\ \bibnamefont {Divall}}, \bibinfo {author} {\bibfnamefont
     {G.}~\bibnamefont {Doucas}}, \bibinfo {author} {\bibfnamefont
     {K.}~\bibnamefont {Ertel}}, \bibinfo {author} {\bibfnamefont
     {F.}~\bibnamefont {Fiuza}}, \bibinfo {author} {\bibfnamefont
     {R.}~\bibnamefont {Fonseca}}, \bibinfo {author} {\bibfnamefont
     {P.}~\bibnamefont {Foster}}, \bibinfo {author} {\bibfnamefont {S.~J.}\
     \bibnamefont {Hawkes}}, \bibinfo {author} {\bibfnamefont {C.~J.}\
     \bibnamefont {Hooker}}, \bibinfo {author} {\bibfnamefont {K.}~\bibnamefont
     {Krushelnick}}, \bibinfo {author} {\bibfnamefont {W.~B.}\ \bibnamefont
     {Mori}}, \bibinfo {author} {\bibfnamefont {C.~A.~J.}\ \bibnamefont {Palmer}},
     \bibinfo {author} {\bibfnamefont {K.~T.}\ \bibnamefont {Phuoc}}, \bibinfo
     {author} {\bibfnamefont {P.~P.}\ \bibnamefont {Rajeev}}, \bibinfo {author}
     {\bibfnamefont {J.}~\bibnamefont {Schreiber}}, \bibinfo {author}
     {\bibfnamefont {M.~J.~V.}\ \bibnamefont {Streeter}}, \bibinfo {author}
     {\bibfnamefont {D.}~\bibnamefont {Urner}}, \bibinfo {author} {\bibfnamefont
     {J.}~\bibnamefont {Vieira}}, \bibinfo {author} {\bibfnamefont {L.~O.}\
     \bibnamefont {Silva}}, \ and\ \bibinfo {author} {\bibfnamefont
     {Z.}~\bibnamefont {Najmudin}},\ }\href {\doibase
     10.1103/PhysRevLett.103.035002} {\bibfield  {journal} {\bibinfo  {journal}
     {Physical Review Letters}\ }\textbf {\bibinfo {volume} {103}},\ \bibinfo
     {pages} {035002} (\bibinfo {year} {2009})}\BibitemShut {NoStop}%
   \bibitem [{\citenamefont {Clayton}\ \emph {et~al.}(2010)\citenamefont
     {Clayton}, \citenamefont {Ralph}, \citenamefont {Albert}, \citenamefont
     {Fonseca}, \citenamefont {Glenzer}, \citenamefont {Joshi}, \citenamefont
     {Lu}, \citenamefont {Marsh}, \citenamefont {Martins}, \citenamefont {Mori},
     \citenamefont {Pak}, \citenamefont {Tsung}, \citenamefont {Pollock},
     \citenamefont {Ross}, \citenamefont {Silva},\ and\ \citenamefont
     {Froula}}]{Clayton2010PRL}%
     \BibitemOpen
     \bibfield  {author} {\bibinfo {author} {\bibfnamefont {C.~E.}\ \bibnamefont
     {Clayton}}, \bibinfo {author} {\bibfnamefont {J.~E.}\ \bibnamefont {Ralph}},
     \bibinfo {author} {\bibfnamefont {F.}~\bibnamefont {Albert}}, \bibinfo
     {author} {\bibfnamefont {R.~A.}\ \bibnamefont {Fonseca}}, \bibinfo {author}
     {\bibfnamefont {S.~H.}\ \bibnamefont {Glenzer}}, \bibinfo {author}
     {\bibfnamefont {C.}~\bibnamefont {Joshi}}, \bibinfo {author} {\bibfnamefont
     {W.}~\bibnamefont {Lu}}, \bibinfo {author} {\bibfnamefont {K.~A.}\
     \bibnamefont {Marsh}}, \bibinfo {author} {\bibfnamefont {S.~F.}\ \bibnamefont
     {Martins}}, \bibinfo {author} {\bibfnamefont {W.~B.}\ \bibnamefont {Mori}},
     \bibinfo {author} {\bibfnamefont {A.}~\bibnamefont {Pak}}, \bibinfo {author}
     {\bibfnamefont {F.~S.}\ \bibnamefont {Tsung}}, \bibinfo {author}
     {\bibfnamefont {B.~B.}\ \bibnamefont {Pollock}}, \bibinfo {author}
     {\bibfnamefont {J.~S.}\ \bibnamefont {Ross}}, \bibinfo {author}
     {\bibfnamefont {L.~O.}\ \bibnamefont {Silva}}, \ and\ \bibinfo {author}
     {\bibfnamefont {D.~H.}\ \bibnamefont {Froula}},\ }\href {\doibase
     10.1103/PhysRevLett.105.105003} {\bibfield  {journal} {\bibinfo  {journal}
     {Physical Review Letters}\ }\textbf {\bibinfo {volume} {105}},\ \bibinfo
     {pages} {105003} (\bibinfo {year} {2010})}\BibitemShut {NoStop}%
   \bibitem [{\citenamefont {Wang}\ \emph {et~al.}(2013)\citenamefont {Wang},
     \citenamefont {Zgadzaj}, \citenamefont {Fazel}, \citenamefont {Li},
     \citenamefont {Yi}, \citenamefont {Zhang}, \citenamefont {Henderson},
     \citenamefont {Chang}, \citenamefont {Korzekwa}, \citenamefont {Tsai},
     \citenamefont {Pai}, \citenamefont {Quevedo}, \citenamefont {Dyer},
     \citenamefont {Gaul}, \citenamefont {Martinez}, \citenamefont {Bernstein},
     \citenamefont {Borger}, \citenamefont {Spinks}, \citenamefont {Donovan},
     \citenamefont {Khudik}, \citenamefont {Shvets}, \citenamefont {Ditmire},\
     and\ \citenamefont {Downer}}]{Wang2013NC}%
     \BibitemOpen
     \bibfield  {author} {\bibinfo {author} {\bibfnamefont {X.}~\bibnamefont
     {Wang}}, \bibinfo {author} {\bibfnamefont {R.}~\bibnamefont {Zgadzaj}},
     \bibinfo {author} {\bibfnamefont {N.}~\bibnamefont {Fazel}}, \bibinfo
     {author} {\bibfnamefont {Z.}~\bibnamefont {Li}}, \bibinfo {author}
     {\bibfnamefont {S.~A.}\ \bibnamefont {Yi}}, \bibinfo {author} {\bibfnamefont
     {X.}~\bibnamefont {Zhang}}, \bibinfo {author} {\bibfnamefont
     {W.}~\bibnamefont {Henderson}}, \bibinfo {author} {\bibfnamefont {Y.-Y.}\
     \bibnamefont {Chang}}, \bibinfo {author} {\bibfnamefont {R.}~\bibnamefont
     {Korzekwa}}, \bibinfo {author} {\bibfnamefont {H.-E.}\ \bibnamefont {Tsai}},
     \bibinfo {author} {\bibfnamefont {C.-H.}\ \bibnamefont {Pai}}, \bibinfo
     {author} {\bibfnamefont {H.}~\bibnamefont {Quevedo}}, \bibinfo {author}
     {\bibfnamefont {G.}~\bibnamefont {Dyer}}, \bibinfo {author} {\bibfnamefont
     {E.}~\bibnamefont {Gaul}}, \bibinfo {author} {\bibfnamefont {M.}~\bibnamefont
     {Martinez}}, \bibinfo {author} {\bibfnamefont {A.~C.}\ \bibnamefont
     {Bernstein}}, \bibinfo {author} {\bibfnamefont {T.}~\bibnamefont {Borger}},
     \bibinfo {author} {\bibfnamefont {M.}~\bibnamefont {Spinks}}, \bibinfo
     {author} {\bibfnamefont {M.}~\bibnamefont {Donovan}}, \bibinfo {author}
     {\bibfnamefont {V.}~\bibnamefont {Khudik}}, \bibinfo {author} {\bibfnamefont
     {G.}~\bibnamefont {Shvets}}, \bibinfo {author} {\bibfnamefont
     {T.}~\bibnamefont {Ditmire}}, \ and\ \bibinfo {author} {\bibfnamefont
     {M.~C.}\ \bibnamefont {Downer}},\ }\href {\doibase 10.1038/ncomms2988}
     {\bibfield  {journal} {\bibinfo  {journal} {Nature Communications}\ }\textbf
     {\bibinfo {volume} {4}},\ \bibinfo {pages} {1988} (\bibinfo {year}
     {2013})}\BibitemShut {NoStop}%
   \bibitem [{\citenamefont {Leemans}\ \emph {et~al.}(2014)\citenamefont
     {Leemans}, \citenamefont {Gonsalves}, \citenamefont {Mao}, \citenamefont
     {Nakamura}, \citenamefont {Benedetti}, \citenamefont {Schroeder},
     \citenamefont {Toth}, \citenamefont {Daniels}, \citenamefont {Mittelberger},
     \citenamefont {Bulanov}, \citenamefont {Vay}, \citenamefont {Geddes},\ and\
     \citenamefont {Esarey}}]{Leemans2014PRL}%
     \BibitemOpen
     \bibfield  {author} {\bibinfo {author} {\bibfnamefont {W.~P.}\ \bibnamefont
     {Leemans}}, \bibinfo {author} {\bibfnamefont {A.~J.}\ \bibnamefont
     {Gonsalves}}, \bibinfo {author} {\bibfnamefont {H.-S.}\ \bibnamefont {Mao}},
     \bibinfo {author} {\bibfnamefont {K.}~\bibnamefont {Nakamura}}, \bibinfo
     {author} {\bibfnamefont {C.}~\bibnamefont {Benedetti}}, \bibinfo {author}
     {\bibfnamefont {C.~B.}\ \bibnamefont {Schroeder}}, \bibinfo {author}
     {\bibfnamefont {C.}~\bibnamefont {Toth}}, \bibinfo {author} {\bibfnamefont
     {J.}~\bibnamefont {Daniels}}, \bibinfo {author} {\bibfnamefont {D.~E.}\
     \bibnamefont {Mittelberger}}, \bibinfo {author} {\bibfnamefont {S.~S.}\
     \bibnamefont {Bulanov}}, \bibinfo {author} {\bibfnamefont {J.-L.}\
     \bibnamefont {Vay}}, \bibinfo {author} {\bibfnamefont {C.~G.~R.}\
     \bibnamefont {Geddes}}, \ and\ \bibinfo {author} {\bibfnamefont
     {E.}~\bibnamefont {Esarey}},\ }\href {\doibase
     10.1103/PhysRevLett.113.245002} {\bibfield  {journal} {\bibinfo  {journal}
     {Physical Review Letters}\ }\textbf {\bibinfo {volume} {113}},\ \bibinfo
     {pages} {245002} (\bibinfo {year} {2014})}\BibitemShut {NoStop}%
   \bibitem [{\citenamefont {Gonsalves}\ \emph {et~al.}(2019)\citenamefont
     {Gonsalves}, \citenamefont {Nakamura}, \citenamefont {Daniels}, \citenamefont
     {Benedetti}, \citenamefont {Pieronek}, \citenamefont {de~Raadt},
     \citenamefont {Steinke}, \citenamefont {Bin}, \citenamefont {Bulanov},
     \citenamefont {van Tilborg}, \citenamefont {Geddes}, \citenamefont
     {Schroeder}, \citenamefont {T{\'{o}}th}, \citenamefont {Esarey},
     \citenamefont {Swanson}, \citenamefont {Fan-Chiang}, \citenamefont
     {Bagdasarov}, \citenamefont {Bobrova}, \citenamefont {Gasilov}, \citenamefont
     {Korn}, \citenamefont {Sasorov},\ and\ \citenamefont
     {Leemans}}]{Gonsalves2019PRL}%
     \BibitemOpen
     \bibfield  {author} {\bibinfo {author} {\bibfnamefont {A.~J.}\ \bibnamefont
     {Gonsalves}}, \bibinfo {author} {\bibfnamefont {K.}~\bibnamefont {Nakamura}},
     \bibinfo {author} {\bibfnamefont {J.}~\bibnamefont {Daniels}}, \bibinfo
     {author} {\bibfnamefont {C.}~\bibnamefont {Benedetti}}, \bibinfo {author}
     {\bibfnamefont {C.}~\bibnamefont {Pieronek}}, \bibinfo {author}
     {\bibfnamefont {T.~C.~H.}\ \bibnamefont {de~Raadt}}, \bibinfo {author}
     {\bibfnamefont {S.}~\bibnamefont {Steinke}}, \bibinfo {author} {\bibfnamefont
     {J.~H.}\ \bibnamefont {Bin}}, \bibinfo {author} {\bibfnamefont {S.~S.}\
     \bibnamefont {Bulanov}}, \bibinfo {author} {\bibfnamefont {J.}~\bibnamefont
     {van Tilborg}}, \bibinfo {author} {\bibfnamefont {C.~G.~R.}\ \bibnamefont
     {Geddes}}, \bibinfo {author} {\bibfnamefont {C.~B.}\ \bibnamefont
     {Schroeder}}, \bibinfo {author} {\bibfnamefont {C.}~\bibnamefont
     {T{\'{o}}th}}, \bibinfo {author} {\bibfnamefont {E.}~\bibnamefont {Esarey}},
     \bibinfo {author} {\bibfnamefont {K.}~\bibnamefont {Swanson}}, \bibinfo
     {author} {\bibfnamefont {L.}~\bibnamefont {Fan-Chiang}}, \bibinfo {author}
     {\bibfnamefont {G.}~\bibnamefont {Bagdasarov}}, \bibinfo {author}
     {\bibfnamefont {N.}~\bibnamefont {Bobrova}}, \bibinfo {author} {\bibfnamefont
     {V.}~\bibnamefont {Gasilov}}, \bibinfo {author} {\bibfnamefont
     {G.}~\bibnamefont {Korn}}, \bibinfo {author} {\bibfnamefont {P.}~\bibnamefont
     {Sasorov}}, \ and\ \bibinfo {author} {\bibfnamefont {W.~P.}\ \bibnamefont
     {Leemans}},\ }\href {\doibase 10.1103/PhysRevLett.122.084801} {\bibfield
     {journal} {\bibinfo  {journal} {Physical Review Letters}\ }\textbf {\bibinfo
     {volume} {122}},\ \bibinfo {pages} {084801} (\bibinfo {year}
     {2019})}\BibitemShut {NoStop}%
   \bibitem [{\citenamefont {Litos}\ \emph {et~al.}(2014)\citenamefont {Litos},
     \citenamefont {Adli}, \citenamefont {An}, \citenamefont {Clarke},
     \citenamefont {Clayton}, \citenamefont {Corde}, \citenamefont {Delahaye},
     \citenamefont {England}, \citenamefont {Fisher}, \citenamefont {Frederico},
     \citenamefont {Gessner}, \citenamefont {Green}, \citenamefont {Hogan},
     \citenamefont {Joshi}, \citenamefont {Lu}, \citenamefont {Marsh},
     \citenamefont {Mori}, \citenamefont {Muggli}, \citenamefont
     {Vafaei-Najafabadi}, \citenamefont {Walz}, \citenamefont {White},
     \citenamefont {Wu}, \citenamefont {Yakimenko},\ and\ \citenamefont
     {Yocky}}]{Litos2014N}%
     \BibitemOpen
     \bibfield  {author} {\bibinfo {author} {\bibfnamefont {M.}~\bibnamefont
     {Litos}}, \bibinfo {author} {\bibfnamefont {E.}~\bibnamefont {Adli}},
     \bibinfo {author} {\bibfnamefont {W.}~\bibnamefont {An}}, \bibinfo {author}
     {\bibfnamefont {C.~I.}\ \bibnamefont {Clarke}}, \bibinfo {author}
     {\bibfnamefont {C.~E.}\ \bibnamefont {Clayton}}, \bibinfo {author}
     {\bibfnamefont {S.}~\bibnamefont {Corde}}, \bibinfo {author} {\bibfnamefont
     {J.~P.}\ \bibnamefont {Delahaye}}, \bibinfo {author} {\bibfnamefont {R.~J.}\
     \bibnamefont {England}}, \bibinfo {author} {\bibfnamefont {A.~S.}\
     \bibnamefont {Fisher}}, \bibinfo {author} {\bibfnamefont {J.}~\bibnamefont
     {Frederico}}, \bibinfo {author} {\bibfnamefont {S.}~\bibnamefont {Gessner}},
     \bibinfo {author} {\bibfnamefont {S.~Z.}\ \bibnamefont {Green}}, \bibinfo
     {author} {\bibfnamefont {M.~J.}\ \bibnamefont {Hogan}}, \bibinfo {author}
     {\bibfnamefont {C.}~\bibnamefont {Joshi}}, \bibinfo {author} {\bibfnamefont
     {W.}~\bibnamefont {Lu}}, \bibinfo {author} {\bibfnamefont {K.~A.}\
     \bibnamefont {Marsh}}, \bibinfo {author} {\bibfnamefont {W.~B.}\ \bibnamefont
     {Mori}}, \bibinfo {author} {\bibfnamefont {P.}~\bibnamefont {Muggli}},
     \bibinfo {author} {\bibfnamefont {N.}~\bibnamefont {Vafaei-Najafabadi}},
     \bibinfo {author} {\bibfnamefont {D.}~\bibnamefont {Walz}}, \bibinfo {author}
     {\bibfnamefont {G.}~\bibnamefont {White}}, \bibinfo {author} {\bibfnamefont
     {Z.}~\bibnamefont {Wu}}, \bibinfo {author} {\bibfnamefont {V.}~\bibnamefont
     {Yakimenko}}, \ and\ \bibinfo {author} {\bibfnamefont {G.}~\bibnamefont
     {Yocky}},\ }\href {\doibase 10.1038/nature13882} {\bibfield  {journal}
     {\bibinfo  {journal} {Nature}\ }\textbf {\bibinfo {volume} {515}},\ \bibinfo
     {pages} {92} (\bibinfo {year} {2014})}\BibitemShut {NoStop}%
   \bibitem [{\citenamefont {Esarey}\ \emph {et~al.}(2009)\citenamefont {Esarey},
     \citenamefont {Schroeder},\ and\ \citenamefont {Leemans}}]{Esarey2009RMP}%
     \BibitemOpen
     \bibfield  {author} {\bibinfo {author} {\bibfnamefont {E.}~\bibnamefont
     {Esarey}}, \bibinfo {author} {\bibfnamefont {C.~B.}\ \bibnamefont
     {Schroeder}}, \ and\ \bibinfo {author} {\bibfnamefont {W.~P.}\ \bibnamefont
     {Leemans}},\ }\href {\doibase 10.1103/RevModPhys.81.1229} {\bibfield
     {journal} {\bibinfo  {journal} {Reviews of Modern Physics}\ }\textbf
     {\bibinfo {volume} {81}},\ \bibinfo {pages} {1229} (\bibinfo {year}
     {2009})}\BibitemShut {NoStop}%
   \bibitem [{\citenamefont {Shiraishi}\ \emph {et~al.}(2013)\citenamefont
     {Shiraishi}, \citenamefont {Benedetti}, \citenamefont {Gonsalves},
     \citenamefont {Nakamura}, \citenamefont {Shaw}, \citenamefont {Sokollik},
     \citenamefont {van Tilborg}, \citenamefont {Geddes}, \citenamefont
     {Schroeder}, \citenamefont {T{\'{o}}th}, \citenamefont {Esarey},\ and\
     \citenamefont {Leemans}}]{Shiraishi2013POP}%
     \BibitemOpen
     \bibfield  {author} {\bibinfo {author} {\bibfnamefont {S.}~\bibnamefont
     {Shiraishi}}, \bibinfo {author} {\bibfnamefont {C.}~\bibnamefont
     {Benedetti}}, \bibinfo {author} {\bibfnamefont {A.~J.}\ \bibnamefont
     {Gonsalves}}, \bibinfo {author} {\bibfnamefont {K.}~\bibnamefont {Nakamura}},
     \bibinfo {author} {\bibfnamefont {B.~H.}\ \bibnamefont {Shaw}}, \bibinfo
     {author} {\bibfnamefont {T.}~\bibnamefont {Sokollik}}, \bibinfo {author}
     {\bibfnamefont {J.}~\bibnamefont {van Tilborg}}, \bibinfo {author}
     {\bibfnamefont {C.~G.~R.}\ \bibnamefont {Geddes}}, \bibinfo {author}
     {\bibfnamefont {C.~B.}\ \bibnamefont {Schroeder}}, \bibinfo {author}
     {\bibfnamefont {C.}~\bibnamefont {T{\'{o}}th}}, \bibinfo {author}
     {\bibfnamefont {E.}~\bibnamefont {Esarey}}, \ and\ \bibinfo {author}
     {\bibfnamefont {W.~P.}\ \bibnamefont {Leemans}},\ }\href {\doibase
     10.1063/1.4810802} {\bibfield  {journal} {\bibinfo  {journal} {Physics of
     Plasmas}\ }\textbf {\bibinfo {volume} {20}},\ \bibinfo {pages} {063103}
     (\bibinfo {year} {2013})}\BibitemShut {NoStop}%
   \bibitem [{\citenamefont {Vieira}\ \emph {et~al.}(2012)\citenamefont {Vieira},
     \citenamefont {Martins}, \citenamefont {Fi{\'{u}}za}, \citenamefont {Huang},
     \citenamefont {Mori}, \citenamefont {Mangles}, \citenamefont {Kneip},
     \citenamefont {Nagel}, \citenamefont {Najmudin},\ and\ \citenamefont
     {Silva}}]{Vieira2012PPCF}%
     \BibitemOpen
     \bibfield  {author} {\bibinfo {author} {\bibfnamefont {J.}~\bibnamefont
     {Vieira}}, \bibinfo {author} {\bibfnamefont {S.~F.}\ \bibnamefont {Martins}},
     \bibinfo {author} {\bibfnamefont {F.}~\bibnamefont {Fi{\'{u}}za}}, \bibinfo
     {author} {\bibfnamefont {C.~K.}\ \bibnamefont {Huang}}, \bibinfo {author}
     {\bibfnamefont {W.~B.}\ \bibnamefont {Mori}}, \bibinfo {author}
     {\bibfnamefont {S.~P.~D.}\ \bibnamefont {Mangles}}, \bibinfo {author}
     {\bibfnamefont {S.}~\bibnamefont {Kneip}}, \bibinfo {author} {\bibfnamefont
     {S.}~\bibnamefont {Nagel}}, \bibinfo {author} {\bibfnamefont
     {Z.}~\bibnamefont {Najmudin}}, \ and\ \bibinfo {author} {\bibfnamefont
     {L.~O.}\ \bibnamefont {Silva}},\ }\href {\doibase
     10.1088/0741-3335/54/5/055010} {\bibfield  {journal} {\bibinfo  {journal}
     {Plasma Physics and Controlled Fusion}\ }\textbf {\bibinfo {volume} {54}},\
     \bibinfo {pages} {55010} (\bibinfo {year} {2012})}\BibitemShut {NoStop}%
   \bibitem [{\citenamefont {Katsouleas}\ \emph {et~al.}(1987)\citenamefont
     {Katsouleas}, \citenamefont {Wilks}, \citenamefont {Chen}, \citenamefont
     {Dawson},\ and\ \citenamefont {Su}}]{Katsouleas1987}%
     \BibitemOpen
     \bibfield  {author} {\bibinfo {author} {\bibfnamefont {T.}~\bibnamefont
     {Katsouleas}}, \bibinfo {author} {\bibfnamefont {S.}~\bibnamefont {Wilks}},
     \bibinfo {author} {\bibfnamefont {P.}~\bibnamefont {Chen}}, \bibinfo {author}
     {\bibfnamefont {J.~M.}\ \bibnamefont {Dawson}}, \ and\ \bibinfo {author}
     {\bibfnamefont {J.~J.}\ \bibnamefont {Su}},\ }\href@noop {} {\bibfield
     {journal} {\bibinfo  {journal} {Particle Accelerators}\ }\textbf {\bibinfo
     {volume} {22}},\ \bibinfo {pages} {81} (\bibinfo {year} {1987})}\BibitemShut
     {NoStop}%
   \bibitem [{\citenamefont {Gordienko}\ and\ \citenamefont
     {Pukhov}(2005)}]{Gordienko2005POP}%
     \BibitemOpen
     \bibfield  {author} {\bibinfo {author} {\bibfnamefont {S.}~\bibnamefont
     {Gordienko}}\ and\ \bibinfo {author} {\bibfnamefont {A.}~\bibnamefont
     {Pukhov}},\ }\href {\doibase 10.1063/1.1884126} {\bibfield  {journal}
     {\bibinfo  {journal} {Physics of Plasmas}\ }\textbf {\bibinfo {volume}
     {12}},\ \bibinfo {pages} {043109} (\bibinfo {year} {2005})}\BibitemShut
     {NoStop}%
   \bibitem [{\citenamefont {Lu}\ \emph {et~al.}(2007)\citenamefont {Lu},
     \citenamefont {Tzoufras}, \citenamefont {Joshi}, \citenamefont {Tsung},
     \citenamefont {Mori}, \citenamefont {Vieira}, \citenamefont {Fonseca},\ and\
     \citenamefont {Silva}}]{Lu2007PRSTAB}%
     \BibitemOpen
     \bibfield  {author} {\bibinfo {author} {\bibfnamefont {W.}~\bibnamefont
     {Lu}}, \bibinfo {author} {\bibfnamefont {M.}~\bibnamefont {Tzoufras}},
     \bibinfo {author} {\bibfnamefont {C.}~\bibnamefont {Joshi}}, \bibinfo
     {author} {\bibfnamefont {F.}~\bibnamefont {Tsung}}, \bibinfo {author}
     {\bibfnamefont {W.}~\bibnamefont {Mori}}, \bibinfo {author} {\bibfnamefont
     {J.}~\bibnamefont {Vieira}}, \bibinfo {author} {\bibfnamefont
     {R.}~\bibnamefont {Fonseca}}, \ and\ \bibinfo {author} {\bibfnamefont
     {L.}~\bibnamefont {Silva}},\ }\href {\doibase 10.1103/PhysRevSTAB.10.061301}
     {\bibfield  {journal} {\bibinfo  {journal} {Physical Review Special Topics -
     Accelerators and Beams}\ }\textbf {\bibinfo {volume} {10}},\ \bibinfo {pages}
     {061301} (\bibinfo {year} {2007})}\BibitemShut {NoStop}%
   \bibitem [{\citenamefont {Tzoufras}\ \emph {et~al.}(2008)\citenamefont
     {Tzoufras}, \citenamefont {Lu}, \citenamefont {Tsung}, \citenamefont {Huang},
     \citenamefont {Mori}, \citenamefont {Katsouleas}, \citenamefont {Vieira},
     \citenamefont {Fonseca},\ and\ \citenamefont {Silva}}]{Tzoufras2008PRL}%
     \BibitemOpen
     \bibfield  {author} {\bibinfo {author} {\bibfnamefont {M.}~\bibnamefont
     {Tzoufras}}, \bibinfo {author} {\bibfnamefont {W.}~\bibnamefont {Lu}},
     \bibinfo {author} {\bibfnamefont {F.~S.}\ \bibnamefont {Tsung}}, \bibinfo
     {author} {\bibfnamefont {C.}~\bibnamefont {Huang}}, \bibinfo {author}
     {\bibfnamefont {W.~B.}\ \bibnamefont {Mori}}, \bibinfo {author}
     {\bibfnamefont {T.}~\bibnamefont {Katsouleas}}, \bibinfo {author}
     {\bibfnamefont {J.}~\bibnamefont {Vieira}}, \bibinfo {author} {\bibfnamefont
     {R.~A.}\ \bibnamefont {Fonseca}}, \ and\ \bibinfo {author} {\bibfnamefont
     {L.~O.}\ \bibnamefont {Silva}},\ }\href {\doibase
     10.1103/PhysRevLett.101.145002} {\bibfield  {journal} {\bibinfo  {journal}
     {Physical Review Letters}\ }\textbf {\bibinfo {volume} {101}},\ \bibinfo
     {pages} {145002} (\bibinfo {year} {2008})}\BibitemShut {NoStop}%
   \bibitem [{\citenamefont {Bulanov}\ \emph {et~al.}(1992)\citenamefont
     {Bulanov}, \citenamefont {Inovenkov}, \citenamefont {Kirsanov}, \citenamefont
     {Naumova},\ and\ \citenamefont {Sakharov}}]{Bulanov1992PoFB}%
     \BibitemOpen
     \bibfield  {author} {\bibinfo {author} {\bibfnamefont {S.~V.}\ \bibnamefont
     {Bulanov}}, \bibinfo {author} {\bibfnamefont {I.~N.}\ \bibnamefont
     {Inovenkov}}, \bibinfo {author} {\bibfnamefont {V.~I.}\ \bibnamefont
     {Kirsanov}}, \bibinfo {author} {\bibfnamefont {N.~M.}\ \bibnamefont
     {Naumova}}, \ and\ \bibinfo {author} {\bibfnamefont {A.~S.}\ \bibnamefont
     {Sakharov}},\ }\href {\doibase 10.1063/1.860046} {\bibfield  {journal}
     {\bibinfo  {journal} {Physics of Fluids B: Plasma Physics}\ }\textbf
     {\bibinfo {volume} {4}},\ \bibinfo {pages} {1935} (\bibinfo {year}
     {1992})}\BibitemShut {NoStop}%
   \bibitem [{\citenamefont {Oliveira~e Silva}\ and\ \citenamefont
     {Mendon{\c{c}}a}(1998)}]{Silva1998PRE}%
     \BibitemOpen
     \bibfield  {author} {\bibinfo {author} {\bibfnamefont {L.}~\bibnamefont
     {Oliveira~e Silva}}\ and\ \bibinfo {author} {\bibfnamefont {J.~T.}\
     \bibnamefont {Mendon{\c{c}}a}},\ }\href {\doibase 10.1103/PhysRevE.57.3423}
     {\bibfield  {journal} {\bibinfo  {journal} {Physical Review E}\ }\textbf
     {\bibinfo {volume} {57}},\ \bibinfo {pages} {3423} (\bibinfo {year}
     {1998})}\BibitemShut {NoStop}%
   \bibitem [{\citenamefont {Hussein}\ \emph {et~al.}(2019)\citenamefont
     {Hussein}, \citenamefont {Senabulya}, \citenamefont {Ma}, \citenamefont
     {Streeter}, \citenamefont {Kettle}, \citenamefont {Dann}, \citenamefont
     {Albert}, \citenamefont {Bourgeois}, \citenamefont {Cipiccia}, \citenamefont
     {Cole}, \citenamefont {Finlay}, \citenamefont {Gerstmayr}, \citenamefont
     {Gonz{\'{a}}lez}, \citenamefont {Higginbotham}, \citenamefont {Jaroszynski},
     \citenamefont {Falk}, \citenamefont {Krushelnick}, \citenamefont {Lemos},
     \citenamefont {Lopes}, \citenamefont {Lumsdon}, \citenamefont {Lundh},
     \citenamefont {Mangles}, \citenamefont {Najmudin}, \citenamefont {Rajeev},
     \citenamefont {Schlep{\"{u}}tz}, \citenamefont {Shahzad}, \citenamefont
     {Smid}, \citenamefont {Spesyvtsev}, \citenamefont {Symes}, \citenamefont
     {Vieux}, \citenamefont {Willingale}, \citenamefont {Wood}, \citenamefont
     {Shahani},\ and\ \citenamefont {Thomas}}]{Hussein2019SR}%
     \BibitemOpen
     \bibfield  {author} {\bibinfo {author} {\bibfnamefont {A.~E.}\ \bibnamefont
     {Hussein}}, \bibinfo {author} {\bibfnamefont {N.}~\bibnamefont {Senabulya}},
     \bibinfo {author} {\bibfnamefont {Y.}~\bibnamefont {Ma}}, \bibinfo {author}
     {\bibfnamefont {M.~J.~V.}\ \bibnamefont {Streeter}}, \bibinfo {author}
     {\bibfnamefont {B.}~\bibnamefont {Kettle}}, \bibinfo {author} {\bibfnamefont
     {S.~J.~D.}\ \bibnamefont {Dann}}, \bibinfo {author} {\bibfnamefont
     {F.}~\bibnamefont {Albert}}, \bibinfo {author} {\bibfnamefont
     {N.}~\bibnamefont {Bourgeois}}, \bibinfo {author} {\bibfnamefont
     {S.}~\bibnamefont {Cipiccia}}, \bibinfo {author} {\bibfnamefont {J.~M.}\
     \bibnamefont {Cole}}, \bibinfo {author} {\bibfnamefont {O.}~\bibnamefont
     {Finlay}}, \bibinfo {author} {\bibfnamefont {E.}~\bibnamefont {Gerstmayr}},
     \bibinfo {author} {\bibfnamefont {I.~G.}\ \bibnamefont {Gonz{\'{a}}lez}},
     \bibinfo {author} {\bibfnamefont {A.}~\bibnamefont {Higginbotham}}, \bibinfo
     {author} {\bibfnamefont {D.~A.}\ \bibnamefont {Jaroszynski}}, \bibinfo
     {author} {\bibfnamefont {K.}~\bibnamefont {Falk}}, \bibinfo {author}
     {\bibfnamefont {K.}~\bibnamefont {Krushelnick}}, \bibinfo {author}
     {\bibfnamefont {N.}~\bibnamefont {Lemos}}, \bibinfo {author} {\bibfnamefont
     {N.~C.}\ \bibnamefont {Lopes}}, \bibinfo {author} {\bibfnamefont
     {C.}~\bibnamefont {Lumsdon}}, \bibinfo {author} {\bibfnamefont
     {O.}~\bibnamefont {Lundh}}, \bibinfo {author} {\bibfnamefont {S.~P.~D.}\
     \bibnamefont {Mangles}}, \bibinfo {author} {\bibfnamefont {Z.}~\bibnamefont
     {Najmudin}}, \bibinfo {author} {\bibfnamefont {P.~P.}\ \bibnamefont
     {Rajeev}}, \bibinfo {author} {\bibfnamefont {C.~M.}\ \bibnamefont
     {Schlep{\"{u}}tz}}, \bibinfo {author} {\bibfnamefont {M.}~\bibnamefont
     {Shahzad}}, \bibinfo {author} {\bibfnamefont {M.}~\bibnamefont {Smid}},
     \bibinfo {author} {\bibfnamefont {R.}~\bibnamefont {Spesyvtsev}}, \bibinfo
     {author} {\bibfnamefont {D.~R.}\ \bibnamefont {Symes}}, \bibinfo {author}
     {\bibfnamefont {G.}~\bibnamefont {Vieux}}, \bibinfo {author} {\bibfnamefont
     {L.}~\bibnamefont {Willingale}}, \bibinfo {author} {\bibfnamefont {J.~C.}\
     \bibnamefont {Wood}}, \bibinfo {author} {\bibfnamefont {A.~J.}\ \bibnamefont
     {Shahani}}, \ and\ \bibinfo {author} {\bibfnamefont {A.~G.~R.}\ \bibnamefont
     {Thomas}},\ }\href {\doibase 10.1038/s41598-019-39845-4} {\bibfield
     {journal} {\bibinfo  {journal} {Scientific Reports}\ }\textbf {\bibinfo
     {volume} {9}},\ \bibinfo {pages} {3249} (\bibinfo {year} {2019})}\BibitemShut
     {NoStop}%
   \bibitem [{\citenamefont {Matsuoka}\ \emph {et~al.}(2010)\citenamefont
     {Matsuoka}, \citenamefont {McGuffey}, \citenamefont {Cummings}, \citenamefont
     {Horovitz}, \citenamefont {Dollar}, \citenamefont {Chvykov}, \citenamefont
     {Kalintchenko}, \citenamefont {Rousseau}, \citenamefont {Yanovsky},
     \citenamefont {Bulanov}, \citenamefont {Thomas}, \citenamefont {Maksimchuk},\
     and\ \citenamefont {Krushelnick}}]{Matsuoka2010PRL}%
     \BibitemOpen
     \bibfield  {author} {\bibinfo {author} {\bibfnamefont {T.}~\bibnamefont
     {Matsuoka}}, \bibinfo {author} {\bibfnamefont {C.}~\bibnamefont {McGuffey}},
     \bibinfo {author} {\bibfnamefont {P.~G.}\ \bibnamefont {Cummings}}, \bibinfo
     {author} {\bibfnamefont {Y.}~\bibnamefont {Horovitz}}, \bibinfo {author}
     {\bibfnamefont {F.}~\bibnamefont {Dollar}}, \bibinfo {author} {\bibfnamefont
     {V.}~\bibnamefont {Chvykov}}, \bibinfo {author} {\bibfnamefont
     {G.}~\bibnamefont {Kalintchenko}}, \bibinfo {author} {\bibfnamefont
     {P.}~\bibnamefont {Rousseau}}, \bibinfo {author} {\bibfnamefont
     {V.}~\bibnamefont {Yanovsky}}, \bibinfo {author} {\bibfnamefont {S.~S.}\
     \bibnamefont {Bulanov}}, \bibinfo {author} {\bibfnamefont {A.~G.~R.}\
     \bibnamefont {Thomas}}, \bibinfo {author} {\bibfnamefont {A.}~\bibnamefont
     {Maksimchuk}}, \ and\ \bibinfo {author} {\bibfnamefont {K.}~\bibnamefont
     {Krushelnick}},\ }\href {\doibase 10.1103/PhysRevLett.105.034801} {\bibfield
     {journal} {\bibinfo  {journal} {Physical Review Letters}\ }\textbf {\bibinfo
     {volume} {105}},\ \bibinfo {pages} {034801} (\bibinfo {year}
     {2010})}\BibitemShut {NoStop}%
   \bibitem [{\citenamefont {Schreiber}\ \emph {et~al.}(2010)\citenamefont
     {Schreiber}, \citenamefont {Bellei}, \citenamefont {Mangles}, \citenamefont
     {Kamperidis}, \citenamefont {Kneip}, \citenamefont {Nagel}, \citenamefont
     {Palmer}, \citenamefont {Rajeev}, \citenamefont {Streeter},\ and\
     \citenamefont {Najmudin}}]{Schreiber2010PRL}%
     \BibitemOpen
     \bibfield  {author} {\bibinfo {author} {\bibfnamefont {J.}~\bibnamefont
     {Schreiber}}, \bibinfo {author} {\bibfnamefont {C.}~\bibnamefont {Bellei}},
     \bibinfo {author} {\bibfnamefont {S.~P.~D.}\ \bibnamefont {Mangles}},
     \bibinfo {author} {\bibfnamefont {C.}~\bibnamefont {Kamperidis}}, \bibinfo
     {author} {\bibfnamefont {S.}~\bibnamefont {Kneip}}, \bibinfo {author}
     {\bibfnamefont {S.~R.}\ \bibnamefont {Nagel}}, \bibinfo {author}
     {\bibfnamefont {C.~A.~J.}\ \bibnamefont {Palmer}}, \bibinfo {author}
     {\bibfnamefont {P.~P.}\ \bibnamefont {Rajeev}}, \bibinfo {author}
     {\bibfnamefont {M.~J.~V.}\ \bibnamefont {Streeter}}, \ and\ \bibinfo {author}
     {\bibfnamefont {Z.}~\bibnamefont {Najmudin}},\ }\href {\doibase
     10.1103/PhysRevLett.105.235003} {\bibfield  {journal} {\bibinfo  {journal}
     {Physical Review Letters}\ }\textbf {\bibinfo {volume} {105}},\ \bibinfo
     {pages} {235003} (\bibinfo {year} {2010})}\BibitemShut {NoStop}%
   \bibitem [{\citenamefont {Lehe}\ \emph {et~al.}(2016)\citenamefont {Lehe},
     \citenamefont {Kirchen}, \citenamefont {Andriyash}, \citenamefont {Godfrey},\
     and\ \citenamefont {Vay}}]{Lehe2016CPC}%
     \BibitemOpen
     \bibfield  {author} {\bibinfo {author} {\bibfnamefont {R.}~\bibnamefont
     {Lehe}}, \bibinfo {author} {\bibfnamefont {M.}~\bibnamefont {Kirchen}},
     \bibinfo {author} {\bibfnamefont {I.~A.}\ \bibnamefont {Andriyash}}, \bibinfo
     {author} {\bibfnamefont {B.~B.}\ \bibnamefont {Godfrey}}, \ and\ \bibinfo
     {author} {\bibfnamefont {J.-L.}\ \bibnamefont {Vay}},\ }\href {\doibase
     10.1016/j.cpc.2016.02.007} {\bibfield  {journal} {\bibinfo  {journal}
     {Computer Physics Communications}\ }\textbf {\bibinfo {volume} {203}},\
     \bibinfo {pages} {66} (\bibinfo {year} {2016})}\BibitemShut {NoStop}%
   \bibitem [{\citenamefont {Mangles}\ \emph {et~al.}(2012)\citenamefont
     {Mangles}, \citenamefont {Genoud}, \citenamefont {Bloom}, \citenamefont
     {Burza}, \citenamefont {Najmudin}, \citenamefont {Persson}, \citenamefont
     {Svensson}, \citenamefont {Thomas},\ and\ \citenamefont
     {Wahlstr{\"{o}}m}}]{Mangles2012PRSTAB}%
     \BibitemOpen
     \bibfield  {author} {\bibinfo {author} {\bibfnamefont {S.~P.~D.}\
     \bibnamefont {Mangles}}, \bibinfo {author} {\bibfnamefont {G.}~\bibnamefont
     {Genoud}}, \bibinfo {author} {\bibfnamefont {M.~S.}\ \bibnamefont {Bloom}},
     \bibinfo {author} {\bibfnamefont {M.}~\bibnamefont {Burza}}, \bibinfo
     {author} {\bibfnamefont {Z.}~\bibnamefont {Najmudin}}, \bibinfo {author}
     {\bibfnamefont {A.}~\bibnamefont {Persson}}, \bibinfo {author} {\bibfnamefont
     {K.}~\bibnamefont {Svensson}}, \bibinfo {author} {\bibfnamefont {A.~G.~R.}\
     \bibnamefont {Thomas}}, \ and\ \bibinfo {author} {\bibfnamefont {C.-G.}\
     \bibnamefont {Wahlstr{\"{o}}m}},\ }\href {\doibase
     10.1103/PhysRevSTAB.15.011302} {\bibfield  {journal} {\bibinfo  {journal}
     {Physical Review Special Topics - Accelerators and Beams}\ }\textbf {\bibinfo
     {volume} {15}},\ \bibinfo {pages} {11302} (\bibinfo {year}
     {2012})}\BibitemShut {NoStop}%
   \bibitem [{\citenamefont {Rowlands-Rees}\ \emph {et~al.}(2008)\citenamefont
     {Rowlands-Rees}, \citenamefont {Kamperidis}, \citenamefont {Kneip},
     \citenamefont {Gonsalves}, \citenamefont {Mangles}, \citenamefont
     {Gallacher}, \citenamefont {Brunetti}, \citenamefont {Ibbotson},
     \citenamefont {Murphy}, \citenamefont {Foster}, \citenamefont {Streeter},
     \citenamefont {Budde}, \citenamefont {Norreys}, \citenamefont {Jaroszynski},
     \citenamefont {Krushelnick}, \citenamefont {Najmudin},\ and\ \citenamefont
     {Hooker}}]{RowlandsRees2008PRL}%
     \BibitemOpen
     \bibfield  {author} {\bibinfo {author} {\bibfnamefont {T.~P.}\ \bibnamefont
     {Rowlands-Rees}}, \bibinfo {author} {\bibfnamefont {C.}~\bibnamefont
     {Kamperidis}}, \bibinfo {author} {\bibfnamefont {S.}~\bibnamefont {Kneip}},
     \bibinfo {author} {\bibfnamefont {A.~J.}\ \bibnamefont {Gonsalves}}, \bibinfo
     {author} {\bibfnamefont {S.~P.~D.}\ \bibnamefont {Mangles}}, \bibinfo
     {author} {\bibfnamefont {J.~G.}\ \bibnamefont {Gallacher}}, \bibinfo {author}
     {\bibfnamefont {E.}~\bibnamefont {Brunetti}}, \bibinfo {author}
     {\bibfnamefont {T.}~\bibnamefont {Ibbotson}}, \bibinfo {author}
     {\bibfnamefont {C.~D.}\ \bibnamefont {Murphy}}, \bibinfo {author}
     {\bibfnamefont {P.~S.}\ \bibnamefont {Foster}}, \bibinfo {author}
     {\bibfnamefont {M.~J.~V.}\ \bibnamefont {Streeter}}, \bibinfo {author}
     {\bibfnamefont {F.}~\bibnamefont {Budde}}, \bibinfo {author} {\bibfnamefont
     {P.~A.}\ \bibnamefont {Norreys}}, \bibinfo {author} {\bibfnamefont {D.~A.}\
     \bibnamefont {Jaroszynski}}, \bibinfo {author} {\bibfnamefont
     {K.}~\bibnamefont {Krushelnick}}, \bibinfo {author} {\bibfnamefont
     {Z.}~\bibnamefont {Najmudin}}, \ and\ \bibinfo {author} {\bibfnamefont
     {S.~M.}\ \bibnamefont {Hooker}},\ }\href@noop {} {\bibfield  {journal}
     {\bibinfo  {journal} {Physical Review Letters}\ }\textbf {\bibinfo {volume}
     {100}},\ \bibinfo {pages} {105005} (\bibinfo {year} {2008})}\BibitemShut
     {NoStop}%
   \bibitem [{\citenamefont {Pak}\ \emph {et~al.}(2010)\citenamefont {Pak},
     \citenamefont {Marsh}, \citenamefont {Martins}, \citenamefont {Lu},
     \citenamefont {Mori},\ and\ \citenamefont {Joshi}}]{Pak2010PRL}%
     \BibitemOpen
     \bibfield  {author} {\bibinfo {author} {\bibfnamefont {A.}~\bibnamefont
     {Pak}}, \bibinfo {author} {\bibfnamefont {K.~A.}\ \bibnamefont {Marsh}},
     \bibinfo {author} {\bibfnamefont {S.~F.}\ \bibnamefont {Martins}}, \bibinfo
     {author} {\bibfnamefont {W.}~\bibnamefont {Lu}}, \bibinfo {author}
     {\bibfnamefont {W.~B.}\ \bibnamefont {Mori}}, \ and\ \bibinfo {author}
     {\bibfnamefont {C.}~\bibnamefont {Joshi}},\ }\href {\doibase
     10.1103/PhysRevLett.104.025003} {\bibfield  {journal} {\bibinfo  {journal}
     {Physical Review Letters}\ }\textbf {\bibinfo {volume} {104}},\ \bibinfo
     {pages} {025003} (\bibinfo {year} {2010})}\BibitemShut {NoStop}%
   \bibitem [{\citenamefont {McGuffey}\ \emph {et~al.}(2010)\citenamefont
     {McGuffey}, \citenamefont {Thomas}, \citenamefont {Schumaker}, \citenamefont
     {Matsuoka}, \citenamefont {Chvykov}, \citenamefont {Dollar}, \citenamefont
     {Kalintchenko}, \citenamefont {Yanovsky}, \citenamefont {Maksimchuk},
     \citenamefont {Krushelnick}, \citenamefont {Bychenkov}, \citenamefont
     {Glazyrin},\ and\ \citenamefont {Karpeev}}]{McGuffey2010PRL}%
     \BibitemOpen
     \bibfield  {author} {\bibinfo {author} {\bibfnamefont {C.}~\bibnamefont
     {McGuffey}}, \bibinfo {author} {\bibfnamefont {A.~G.~R.}\ \bibnamefont
     {Thomas}}, \bibinfo {author} {\bibfnamefont {W.}~\bibnamefont {Schumaker}},
     \bibinfo {author} {\bibfnamefont {T.}~\bibnamefont {Matsuoka}}, \bibinfo
     {author} {\bibfnamefont {V.}~\bibnamefont {Chvykov}}, \bibinfo {author}
     {\bibfnamefont {F.~J.}\ \bibnamefont {Dollar}}, \bibinfo {author}
     {\bibfnamefont {G.}~\bibnamefont {Kalintchenko}}, \bibinfo {author}
     {\bibfnamefont {V.}~\bibnamefont {Yanovsky}}, \bibinfo {author}
     {\bibfnamefont {A.}~\bibnamefont {Maksimchuk}}, \bibinfo {author}
     {\bibfnamefont {K.}~\bibnamefont {Krushelnick}}, \bibinfo {author}
     {\bibfnamefont {V.~Y.}\ \bibnamefont {Bychenkov}}, \bibinfo {author}
     {\bibfnamefont {I.~V.}\ \bibnamefont {Glazyrin}}, \ and\ \bibinfo {author}
     {\bibfnamefont {a.~V.}\ \bibnamefont {Karpeev}},\ }\href {\doibase
     10.1103/PhysRevLett.104.025004} {\bibfield  {journal} {\bibinfo  {journal}
     {Physical Review Letters}\ }\textbf {\bibinfo {volume} {104}},\ \bibinfo
     {pages} {025004} (\bibinfo {year} {2010})}\BibitemShut {NoStop}%
   \bibitem [{\citenamefont {Chen}\ \emph {et~al.}(2012)\citenamefont {Chen},
     \citenamefont {Esarey}, \citenamefont {Schroeder}, \citenamefont {Geddes},\
     and\ \citenamefont {Leemans}}]{Chen2012POP}%
     \BibitemOpen
     \bibfield  {author} {\bibinfo {author} {\bibfnamefont {M.}~\bibnamefont
     {Chen}}, \bibinfo {author} {\bibfnamefont {E.}~\bibnamefont {Esarey}},
     \bibinfo {author} {\bibfnamefont {C.~B.}\ \bibnamefont {Schroeder}}, \bibinfo
     {author} {\bibfnamefont {C.~G.~R.}\ \bibnamefont {Geddes}}, \ and\ \bibinfo
     {author} {\bibfnamefont {W.~P.}\ \bibnamefont {Leemans}},\ }\href {\doibase
     10.1063/1.3689922} {\bibfield  {journal} {\bibinfo  {journal} {Physics of
     Plasmas}\ }\textbf {\bibinfo {volume} {19}},\ \bibinfo {pages} {033101}
     (\bibinfo {year} {2012})}\BibitemShut {NoStop}%
   \bibitem [{\citenamefont {Pukhov}\ and\ \citenamefont
     {Kostyukov}(2008)}]{Pukhov2008PRE}%
     \BibitemOpen
     \bibfield  {author} {\bibinfo {author} {\bibfnamefont {A.}~\bibnamefont
     {Pukhov}}\ and\ \bibinfo {author} {\bibfnamefont {I.}~\bibnamefont
     {Kostyukov}},\ }\href {\doibase 10.1103/PhysRevE.77.025401} {\bibfield
     {journal} {\bibinfo  {journal} {Physical Review E}\ }\textbf {\bibinfo
     {volume} {77}},\ \bibinfo {pages} {025401(R)} (\bibinfo {year}
     {2008})}\BibitemShut {NoStop}%
   \bibitem [{\citenamefont {Guillaume}\ \emph {et~al.}(2015)\citenamefont
     {Guillaume}, \citenamefont {D{\"{o}}pp}, \citenamefont {Thaury},
     \citenamefont {Ta~Phuoc}, \citenamefont {Lifschitz}, \citenamefont
     {Grittani}, \citenamefont {Goddet}, \citenamefont {Tafzi}, \citenamefont
     {Chou}, \citenamefont {Veisz},\ and\ \citenamefont
     {Malka}}]{Guillaume2015PRL}%
     \BibitemOpen
     \bibfield  {author} {\bibinfo {author} {\bibfnamefont {E.}~\bibnamefont
     {Guillaume}}, \bibinfo {author} {\bibfnamefont {A.}~\bibnamefont
     {D{\"{o}}pp}}, \bibinfo {author} {\bibfnamefont {C.}~\bibnamefont {Thaury}},
     \bibinfo {author} {\bibfnamefont {K.}~\bibnamefont {Ta~Phuoc}}, \bibinfo
     {author} {\bibfnamefont {A.}~\bibnamefont {Lifschitz}}, \bibinfo {author}
     {\bibfnamefont {G.}~\bibnamefont {Grittani}}, \bibinfo {author}
     {\bibfnamefont {J.~P.}\ \bibnamefont {Goddet}}, \bibinfo {author}
     {\bibfnamefont {A.}~\bibnamefont {Tafzi}}, \bibinfo {author} {\bibfnamefont
     {S.~W.}\ \bibnamefont {Chou}}, \bibinfo {author} {\bibfnamefont
     {L.}~\bibnamefont {Veisz}}, \ and\ \bibinfo {author} {\bibfnamefont
     {V.}~\bibnamefont {Malka}},\ }\href {\doibase 10.1103/PhysRevLett.115.155002}
     {\bibfield  {journal} {\bibinfo  {journal} {Physical Review Letters}\
     }\textbf {\bibinfo {volume} {115}},\ \bibinfo {pages} {155002} (\bibinfo
     {year} {2015})}\BibitemShut {NoStop}%
   \bibitem [{\citenamefont {Ma}\ \emph {et~al.}(2018)\citenamefont {Ma},
     \citenamefont {Seipt}, \citenamefont {Dann}, \citenamefont {Streeter},
     \citenamefont {Palmer}, \citenamefont {Willingale},\ and\ \citenamefont
     {Thomas}}]{Ma2018POP}%
     \BibitemOpen
     \bibfield  {author} {\bibinfo {author} {\bibfnamefont {Y.}~\bibnamefont
     {Ma}}, \bibinfo {author} {\bibfnamefont {D.}~\bibnamefont {Seipt}}, \bibinfo
     {author} {\bibfnamefont {S.~J.~D.}\ \bibnamefont {Dann}}, \bibinfo {author}
     {\bibfnamefont {M.~J.~V.}\ \bibnamefont {Streeter}}, \bibinfo {author}
     {\bibfnamefont {C.~A.~J.}\ \bibnamefont {Palmer}}, \bibinfo {author}
     {\bibfnamefont {L.}~\bibnamefont {Willingale}}, \ and\ \bibinfo {author}
     {\bibfnamefont {A.~G.~R.}\ \bibnamefont {Thomas}},\ }\href {\doibase
     10.1063/1.5054807} {\bibfield  {journal} {\bibinfo  {journal} {Physics of
     Plasmas}\ }\textbf {\bibinfo {volume} {25}},\ \bibinfo {pages} {113105}
     (\bibinfo {year} {2018})}\BibitemShut {NoStop}%
   \bibitem [{\citenamefont {Sadler}\ \emph {et~al.}(2020)\citenamefont {Sadler},
     \citenamefont {Arran}, \citenamefont {Li},\ and\ \citenamefont
     {Flippo}}]{Sadler2020PRAB}%
     \BibitemOpen
     \bibfield  {author} {\bibinfo {author} {\bibfnamefont {J.~D.}\ \bibnamefont
     {Sadler}}, \bibinfo {author} {\bibfnamefont {C.}~\bibnamefont {Arran}},
     \bibinfo {author} {\bibfnamefont {H.}~\bibnamefont {Li}}, \ and\ \bibinfo
     {author} {\bibfnamefont {K.~A.}\ \bibnamefont {Flippo}},\ }\href {\doibase
     10.1103/PhysRevAccelBeams.23.021303} {\bibfield  {journal} {\bibinfo
     {journal} {Physical Review Accelerators and Beams}\ }\textbf {\bibinfo
     {volume} {23}},\ \bibinfo {pages} {021303} (\bibinfo {year}
     {2020})}\BibitemShut {NoStop}%
   \bibitem [{\citenamefont {Debus}\ \emph {et~al.}(2019)\citenamefont {Debus},
     \citenamefont {Pausch}, \citenamefont {Huebl}, \citenamefont {Steiniger},
     \citenamefont {Widera}, \citenamefont {Cowan}, \citenamefont {Schramm},\ and\
     \citenamefont {Bussmann}}]{Debus2019PRX}%
     \BibitemOpen
     \bibfield  {author} {\bibinfo {author} {\bibfnamefont {A.}~\bibnamefont
     {Debus}}, \bibinfo {author} {\bibfnamefont {R.}~\bibnamefont {Pausch}},
     \bibinfo {author} {\bibfnamefont {A.}~\bibnamefont {Huebl}}, \bibinfo
     {author} {\bibfnamefont {K.}~\bibnamefont {Steiniger}}, \bibinfo {author}
     {\bibfnamefont {R.}~\bibnamefont {Widera}}, \bibinfo {author} {\bibfnamefont
     {T.~E.}\ \bibnamefont {Cowan}}, \bibinfo {author} {\bibfnamefont
     {U.}~\bibnamefont {Schramm}}, \ and\ \bibinfo {author} {\bibfnamefont
     {M.}~\bibnamefont {Bussmann}},\ }\href {\doibase 10.1103/PhysRevX.9.031044}
     {\bibfield  {journal} {\bibinfo  {journal} {Physical Review X}\ }\textbf
     {\bibinfo {volume} {9}},\ \bibinfo {pages} {031044} (\bibinfo {year}
     {2019})}\BibitemShut {NoStop}%
   \bibitem [{\citenamefont {Palastro}\ \emph {et~al.}(2020)\citenamefont
     {Palastro}, \citenamefont {Shaw}, \citenamefont {Franke}, \citenamefont
     {Ramsey}, \citenamefont {Simpson},\ and\ \citenamefont
     {Froula}}]{Palastro2020PRL}%
     \BibitemOpen
     \bibfield  {author} {\bibinfo {author} {\bibfnamefont {J.~P.}\ \bibnamefont
     {Palastro}}, \bibinfo {author} {\bibfnamefont {J.~L.}\ \bibnamefont {Shaw}},
     \bibinfo {author} {\bibfnamefont {P.}~\bibnamefont {Franke}}, \bibinfo
     {author} {\bibfnamefont {D.}~\bibnamefont {Ramsey}}, \bibinfo {author}
     {\bibfnamefont {T.~T.}\ \bibnamefont {Simpson}}, \ and\ \bibinfo {author}
     {\bibfnamefont {D.~H.}\ \bibnamefont {Froula}},\ }\href {\doibase
     10.1103/PhysRevLett.124.134802} {\bibfield  {journal} {\bibinfo  {journal}
     {Physical Review Letters}\ }\textbf {\bibinfo {volume} {124}},\ \bibinfo
     {pages} {134802} (\bibinfo {year} {2020})}\BibitemShut {NoStop}%
   \bibitem [{\citenamefont {Caizergues}\ \emph {et~al.}(2020)\citenamefont
     {Caizergues}, \citenamefont {Smartsev}, \citenamefont {Malka},\ and\
     \citenamefont {Thaury}}]{Caizergues2020Nph}%
     \BibitemOpen
     \bibfield  {author} {\bibinfo {author} {\bibfnamefont {C.}~\bibnamefont
     {Caizergues}}, \bibinfo {author} {\bibfnamefont {S.}~\bibnamefont
     {Smartsev}}, \bibinfo {author} {\bibfnamefont {V.}~\bibnamefont {Malka}}, \
     and\ \bibinfo {author} {\bibfnamefont {C.}~\bibnamefont {Thaury}},\ }\href
     {\doibase 10.1038/s41566-020-0657-2} {\bibfield  {journal} {\bibinfo
     {journal} {Nature Photonics}\ }\textbf {\bibinfo {volume} {14}},\ \bibinfo
     {pages} {475} (\bibinfo {year} {2020})}\BibitemShut {NoStop}%
   \bibitem [{\citenamefont {Li}\ \emph {et~al.}(2014)\citenamefont {Li},
     \citenamefont {Liu}, \citenamefont {Wang}, \citenamefont {Zhang},
     \citenamefont {Chen}, \citenamefont {Tian}, \citenamefont {Qi}, \citenamefont
     {Yu}, \citenamefont {Wang}, \citenamefont {Tajima}, \citenamefont {Li},\ and\
     \citenamefont {Xu}}]{Li2014APL}%
     \BibitemOpen
     \bibfield  {author} {\bibinfo {author} {\bibfnamefont {W.}~\bibnamefont
     {Li}}, \bibinfo {author} {\bibfnamefont {J.}~\bibnamefont {Liu}}, \bibinfo
     {author} {\bibfnamefont {W.}~\bibnamefont {Wang}}, \bibinfo {author}
     {\bibfnamefont {Z.}~\bibnamefont {Zhang}}, \bibinfo {author} {\bibfnamefont
     {Q.}~\bibnamefont {Chen}}, \bibinfo {author} {\bibfnamefont {Y.}~\bibnamefont
     {Tian}}, \bibinfo {author} {\bibfnamefont {R.}~\bibnamefont {Qi}}, \bibinfo
     {author} {\bibfnamefont {C.}~\bibnamefont {Yu}}, \bibinfo {author}
     {\bibfnamefont {C.}~\bibnamefont {Wang}}, \bibinfo {author} {\bibfnamefont
     {T.}~\bibnamefont {Tajima}}, \bibinfo {author} {\bibfnamefont
     {R.}~\bibnamefont {Li}}, \ and\ \bibinfo {author} {\bibfnamefont
     {Z.}~\bibnamefont {Xu}},\ }\href {\doibase 10.1063/1.4867536} {\bibfield
     {journal} {\bibinfo  {journal} {Applied Physics Letters}\ }\textbf {\bibinfo
     {volume} {104}},\ \bibinfo {pages} {093510} (\bibinfo {year}
     {2014})}\BibitemShut {NoStop}%
   \bibitem [{\citenamefont {Streeter}\ \emph {et~al.}(2018)\citenamefont
     {Streeter}, \citenamefont {Kneip}, \citenamefont {Bloom}, \citenamefont
     {Bendoyro}, \citenamefont {Chekhlov}, \citenamefont {Dangor}, \citenamefont
     {D{\"{o}}pp}, \citenamefont {Hooker}, \citenamefont {Holloway}, \citenamefont
     {Jiang}, \citenamefont {Lopes}, \citenamefont {Nakamura}, \citenamefont
     {Norreys}, \citenamefont {Palmer}, \citenamefont {Rajeev}, \citenamefont
     {Schreiber}, \citenamefont {Symes}, \citenamefont {Wing}, \citenamefont
     {Mangles},\ and\ \citenamefont {Najmudin}}]{Streeter2018PRL}%
     \BibitemOpen
     \bibfield  {author} {\bibinfo {author} {\bibfnamefont {M.~J.~V.}\
     \bibnamefont {Streeter}}, \bibinfo {author} {\bibfnamefont {S.}~\bibnamefont
     {Kneip}}, \bibinfo {author} {\bibfnamefont {M.~S.}\ \bibnamefont {Bloom}},
     \bibinfo {author} {\bibfnamefont {R.~A.}\ \bibnamefont {Bendoyro}}, \bibinfo
     {author} {\bibfnamefont {O.}~\bibnamefont {Chekhlov}}, \bibinfo {author}
     {\bibfnamefont {A.~E.}\ \bibnamefont {Dangor}}, \bibinfo {author}
     {\bibfnamefont {A.}~\bibnamefont {D{\"{o}}pp}}, \bibinfo {author}
     {\bibfnamefont {C.~J.}\ \bibnamefont {Hooker}}, \bibinfo {author}
     {\bibfnamefont {J.}~\bibnamefont {Holloway}}, \bibinfo {author}
     {\bibfnamefont {J.}~\bibnamefont {Jiang}}, \bibinfo {author} {\bibfnamefont
     {N.~C.}\ \bibnamefont {Lopes}}, \bibinfo {author} {\bibfnamefont
     {H.}~\bibnamefont {Nakamura}}, \bibinfo {author} {\bibfnamefont {P.~A.}\
     \bibnamefont {Norreys}}, \bibinfo {author} {\bibfnamefont {C.~A.~J.}\
     \bibnamefont {Palmer}}, \bibinfo {author} {\bibfnamefont {P.~P.}\
     \bibnamefont {Rajeev}}, \bibinfo {author} {\bibfnamefont {J.}~\bibnamefont
     {Schreiber}}, \bibinfo {author} {\bibfnamefont {D.~R.}\ \bibnamefont
     {Symes}}, \bibinfo {author} {\bibfnamefont {M.}~\bibnamefont {Wing}},
     \bibinfo {author} {\bibfnamefont {S.~P.~D.}\ \bibnamefont {Mangles}}, \ and\
     \bibinfo {author} {\bibfnamefont {Z.}~\bibnamefont {Najmudin}},\ }\href
     {\doibase 10.1103/PhysRevLett.120.254801} {\bibfield  {journal} {\bibinfo
     {journal} {Physical Review Letters}\ }\textbf {\bibinfo {volume} {120}},\
     \bibinfo {pages} {254801} (\bibinfo {year} {2018})}\BibitemShut {NoStop}%
   \bibitem [{\citenamefont {Nie}\ \emph {et~al.}(2018)\citenamefont {Nie},
     \citenamefont {Pai}, \citenamefont {Hua}, \citenamefont {Zhang},
     \citenamefont {Wu}, \citenamefont {Wan}, \citenamefont {Li}, \citenamefont
     {Zhang}, \citenamefont {Cheng}, \citenamefont {Su}, \citenamefont {Liu},
     \citenamefont {Ma}, \citenamefont {Ning}, \citenamefont {He}, \citenamefont
     {Lu}, \citenamefont {Chu}, \citenamefont {Wang}, \citenamefont {Mori},\ and\
     \citenamefont {Joshi}}]{Nie2018Nph}%
     \BibitemOpen
     \bibfield  {author} {\bibinfo {author} {\bibfnamefont {Z.}~\bibnamefont
     {Nie}}, \bibinfo {author} {\bibfnamefont {C.~H.}\ \bibnamefont {Pai}},
     \bibinfo {author} {\bibfnamefont {J.}~\bibnamefont {Hua}}, \bibinfo {author}
     {\bibfnamefont {C.}~\bibnamefont {Zhang}}, \bibinfo {author} {\bibfnamefont
     {Y.}~\bibnamefont {Wu}}, \bibinfo {author} {\bibfnamefont {Y.}~\bibnamefont
     {Wan}}, \bibinfo {author} {\bibfnamefont {F.}~\bibnamefont {Li}}, \bibinfo
     {author} {\bibfnamefont {J.}~\bibnamefont {Zhang}}, \bibinfo {author}
     {\bibfnamefont {Z.}~\bibnamefont {Cheng}}, \bibinfo {author} {\bibfnamefont
     {Q.}~\bibnamefont {Su}}, \bibinfo {author} {\bibfnamefont {S.}~\bibnamefont
     {Liu}}, \bibinfo {author} {\bibfnamefont {Y.}~\bibnamefont {Ma}}, \bibinfo
     {author} {\bibfnamefont {X.}~\bibnamefont {Ning}}, \bibinfo {author}
     {\bibfnamefont {Y.}~\bibnamefont {He}}, \bibinfo {author} {\bibfnamefont
     {W.}~\bibnamefont {Lu}}, \bibinfo {author} {\bibfnamefont {H.~H.}\
     \bibnamefont {Chu}}, \bibinfo {author} {\bibfnamefont {J.}~\bibnamefont
     {Wang}}, \bibinfo {author} {\bibfnamefont {W.~B.}\ \bibnamefont {Mori}}, \
     and\ \bibinfo {author} {\bibfnamefont {C.}~\bibnamefont {Joshi}},\ }\href
     {\doibase 10.1038/s41566-018-0190-8} {\bibfield  {journal} {\bibinfo
     {journal} {Nature Photonics}\ }\textbf {\bibinfo {volume} {12}},\ \bibinfo
     {pages} {489} (\bibinfo {year} {2018})}\BibitemShut {NoStop}%
   \bibitem [{\citenamefont {Nie}\ \emph {et~al.}(2020)\citenamefont {Nie},
     \citenamefont {Pai}, \citenamefont {Zhang}, \citenamefont {Ning},
     \citenamefont {Hua}, \citenamefont {He}, \citenamefont {Wu}, \citenamefont
     {Su}, \citenamefont {Liu}, \citenamefont {Ma}, \citenamefont {Cheng},
     \citenamefont {Lu}, \citenamefont {Chu}, \citenamefont {Wang}, \citenamefont
     {Zhang}, \citenamefont {Mori},\ and\ \citenamefont {Joshi}}]{Nie2020NC}%
     \BibitemOpen
     \bibfield  {author} {\bibinfo {author} {\bibfnamefont {Z.}~\bibnamefont
     {Nie}}, \bibinfo {author} {\bibfnamefont {C.-H.}\ \bibnamefont {Pai}},
     \bibinfo {author} {\bibfnamefont {J.}~\bibnamefont {Zhang}}, \bibinfo
     {author} {\bibfnamefont {X.}~\bibnamefont {Ning}}, \bibinfo {author}
     {\bibfnamefont {J.}~\bibnamefont {Hua}}, \bibinfo {author} {\bibfnamefont
     {Y.}~\bibnamefont {He}}, \bibinfo {author} {\bibfnamefont {Y.}~\bibnamefont
     {Wu}}, \bibinfo {author} {\bibfnamefont {Q.}~\bibnamefont {Su}}, \bibinfo
     {author} {\bibfnamefont {S.}~\bibnamefont {Liu}}, \bibinfo {author}
     {\bibfnamefont {Y.}~\bibnamefont {Ma}}, \bibinfo {author} {\bibfnamefont
     {Z.}~\bibnamefont {Cheng}}, \bibinfo {author} {\bibfnamefont
     {W.}~\bibnamefont {Lu}}, \bibinfo {author} {\bibfnamefont {H.-H.}\
     \bibnamefont {Chu}}, \bibinfo {author} {\bibfnamefont {J.}~\bibnamefont
     {Wang}}, \bibinfo {author} {\bibfnamefont {C.}~\bibnamefont {Zhang}},
     \bibinfo {author} {\bibfnamefont {W.~B.}\ \bibnamefont {Mori}}, \ and\
     \bibinfo {author} {\bibfnamefont {C.}~\bibnamefont {Joshi}},\ }\href
     {\doibase 10.1038/s41467-020-16541-w} {\bibfield  {journal} {\bibinfo
     {journal} {Nature Communications}\ }\textbf {\bibinfo {volume} {11}},\
     \bibinfo {pages} {2787} (\bibinfo {year} {2020})}\BibitemShut {NoStop}%
   \bibitem [{\citenamefont {Streeter}(2022)}]{Streeter2022zenodo}%
     \BibitemOpen
     \bibfield  {author} {\bibinfo {author} {\bibfnamefont {M.~J.~V.}\
     \bibnamefont {Streeter}},\ }\href {\doibase 10.5281/zenodo.7188057} {\
     (\bibinfo {year} {2022}),\ 10.5281/zenodo.7188057}\BibitemShut {NoStop}%
   \end{thebibliography}
\end{document}